\documentstyle[epsfig]{elsart}
\def\ltap{\;\raisebox{-.5ex}{\rlap{$\sim$}} \raisebox{.5ex}{$<$}\;}

\newcommand{\be}{\begin{equation}}  
\newcommand{\ee}{\end{equation}}
\newcommand{\ba}{\begin{eqnarray}}  
\newcommand{\ea}{\end{eqnarray}}

\newcommand{\ewxy}[2]{\setlength{\epsfxsize}{#2}\epsfbox[45 240 420 350]{#1}}

\newcommand{\putfig}[1]{
\vspace{3.0truecm}\ewxy{#1}{120mm}\vspace{2.5truecm}}
\begin{document}

\rightline{Rome Preprint 1157/96}
\rightline{SNS/PH/1996-007}
\rightline{ SWAT/133}
\rightline{ FTUV 96/73 - IFIC 96/82}

\begin{frontmatter}

\title{Light Quenched Hadron Spectrum and Decay Constants on different 
       Lattices}

\author{C.R.~Allton}
\address{Department of Physics, University of Wales Swansea, 
Singleton Park, \\ Swansea SA2 8PP, United Kingdom}
\author{V.~Gim\'enez}
\address{Dep. de Fisica Teorica and IFIC, Univ. de Val\`encia,\\
Dr. Moliner 50, E-46100, Burjassot, Val\`encia, Spain}
\author{L.~Giusti}
\address{Scuola Normale Superiore, P.za dei Cavalieri 7 - I-56100 Pisa Italy\\ 
         INFN Sezione di Pisa I-56100 Pisa Italy}
\author{F.~Rapuano}
\address{Dipartimento di Fisica, Universit\`a di Roma \lq La Sapienza\rq ~and\\
         INFN, Sezione di Roma, P.le A. Moro 2, I-00185 Roma, Italy.}

\begin{abstract} In this paper we study ${\cal O}(2000)$ (quenched)
lattice configurations from the APE collaboration, for different lattice
volumes and for $6.0 \le \beta
\le 6.4$ using both the Wilson and the SW-Clover fermion actions. 
We determine the light hadronic spectrum and meson decay constants and study
the mesonic dispersion relation. We extract the hadronic variable $J$ and the
strange quark mass in the continuum at the next-to-leading order obtaining 
$m_s^{\overline{MS}}(\mu =2 GeV) = 122 \pm 20\; MeV $. A study is
made of their dependence on lattice spacing. We implement a newly
developed technique to extract the inverse
lattice spacing using data at the {\em simulated values}  of the quark mass
(i.e. at masses around the strange quark mass).
\end{abstract}

\end{frontmatter}

\vfill
\centerline{PACS: 11.15.H, 12.38.Gc, 13.30.Eg, 14.20.-c and 14.40.-n}

\newpage
\clearpage 

\section{Introduction}
The lattice technique has proved a very effective theoretical tool to determine
phenomenological quantities such as the mass spectrum and weak decay matrix
elements. Unlike other approaches, it does not (in principle) suffer from
uncontrolled approximations. However, in practice, one is forced to work on a
lattice of (i) finite size, with (ii) a finite lattice spacing, and (iii)
unphysically large masses for the light quarks. 
Also the quenched approximation is often used in lattice studies.
In this work we determine the
light hadronic spectrum and  meson decay constants and we study the effects of
(i), (ii) and (iii), and in particular, we try to  overcome some of the
inherent approximations corresponding to (iii). It is important to study  the
systematics due to (i), (ii) and (iii) before the effects of the quenched 
approximation can be correctly understood.\\
The results reported in this paper are obtained from ${\cal O}(2000)$ (quenched)
lattice configurations from the APE collaboration, for different lattice
volumes and for $6.0 \le \beta \le 6.4$ using both the Wilson action and the  
SW-Clover fermion action.
In this  range of $\beta$ we cannot draw any conclusion about $a$ dependence
of hadron masses and pseudoscalar decay constants. On the other hand we find
that the quenched approximation gives a resonable agreement with the
experimental values.\\
The plan of the paper is as follows. In the next section details of the
simulations are given. Sect.~\ref{new_method} describes the method used
to determine the spectrum and matrix elements from the lattice data. 
The results on hadron masses, meson decay constants, $J$ values, meson
dispersion relations and quark 
masses are then presented in Sect.~\ref{results}, with
a comparison with results from other collaborations. 
In the last section we report our concluding remarks.

\section{Lattice Details}
\label{lat_details}
In the last four years the APE group has been extensively studying lattice QCD
in the quenched approximation.
\setlength{\tabcolsep}{.18pc}
\begin{table}
\label{tab:latparams} 
\caption{Summary of the parameters of the runs analyzed in this work.
T-Sweeps stands for thermalising Montecarlo sweeps performed starting from a 
cold configuration and I-Sweeps stands for the number of sweeps after which 
independent configurations are gathered.}
\begin{tabular}{ccccccccc}
\hline         
&C60&W60&C62a&W62a&C62b&W62b&W64&C64\\
\hline
Ref & \cite{latI}-\cite{latI2} & \cite{latI,latII_a} & \cite{latIII} & \cite{latIII}
    & \cite{latV} & \cite{latII_a,latII_a2} & \cite{latIII} & \cite{latIII} \\
$\beta$&$6.0$&$6.0$&$6.2$ &$6.2$&$6.2$&$6.2$&$6.4$&$6.4$\\
Action & SW & Wil & SW & Wil& SW & Wil & Wil & SW \\
\# Confs&200&320&250&250&200&110&400&400\\
Volume&$18^3\times 64$&$18^3\times 64$&$24^3\times 64$& $24^3\times 64$ 
&$18^3\times 64$&$24^3\times 64$&$24^3\times 64$&$24^3\times 64$\\
T-Sweeps& 3000 & 3000 & 5000 & 5000 & 3000 & 3000 & 7000 & 7000\\
I-Sweeps& 800  & 800  & 2000 & 2000 & 800  & 2400 & 3000 & 3000\\
\hline
$K$&  -   &  -   &0.14144&0.1510&    -   &  -   &0.1488&0.1400\\
   &0.1425&0.1530&0.14184&0.1515& 0.14144&0.1510&0.1492&0.1403\\
   &0.1432&0.1540&0.14224&0.1520& 0.14190&0.1520&0.1496&0.1406\\
   &0.1440&0.1550&0.14264&0.1526& 0.14244&0.1526&0.1500&0.1409\\
\hline
\hline
\end{tabular}
\end{table}
The lattice data used in this study come from ${\cal O}(2000)$ configurations 
generated by the APE collaboration primarily to study weak matrix elements 
such as $f_D$, $f_B$ and $B_K$ and semileptonic decays \cite{latI}-\cite{latV}. 
The data have been obtained from eight sets of configurations with $\beta=6.0$,
$6.2$ and $6.4$.  
The quark propagators have been 
obtained from either the Wilson action or the `improved' SW-Clover
action, by the standard APE inversion procedure\cite{ape57}.  
The hopping parameters   
and the other parameters used in each simulation are 
listed in Table~1.
These $K$ values correspond to quark 
masses roughly in the range of the strange quark mass which are adequate for
the primary goal of the runs  described before.
In the past the APE collaboration has introduced the so called `smearing'
technique \cite{ape57} in order to have a better determination of the lowest
state mass. The APE group has compared data extracted from smeared and
non-smeared propagators and found no real improvement on lattices of a time
extension of 64 sites at $\beta=6.0$ and $\beta=6.2$ \cite{latII_a2}. The data
presented in this work are obtained without any smearing procedure. In fact
the data from W64 and C64 seem to suggest that a smearing procedure would have
increased the reliability of the extraction of the mass of the lowest lying
states.
We plan to increase the time extension of this lattice and check the validity 
of the `local' data presented here.\\
A further comment is due regarding the `thinning' procedure \cite{ape57} 
used to reduce memory occupation. 
In this technique, correlation functions are computed
taking only one point out of three in each space direction. We have checked 
that this procedure does not affect the physical quantities extracted from 
the two point correlation functions. All baryonic data
presented in this paper come from `thinned' propagators while mesonic
correlation functions come from the full lattice.\\  
Note that the simulations have been performed at $\beta$ values of 
$6.0$ or larger. This is to negate
the large systematic error present in lattice data for $\beta \ltap
6.0$ due to lattice artifacts. A clear signal for these systematics is
the lack of scaling amongst different physical quantities (see, for
example, table 2 of \cite{cra} and \cite{cra2}). 

\begin{figure}[htb] 
\putfig{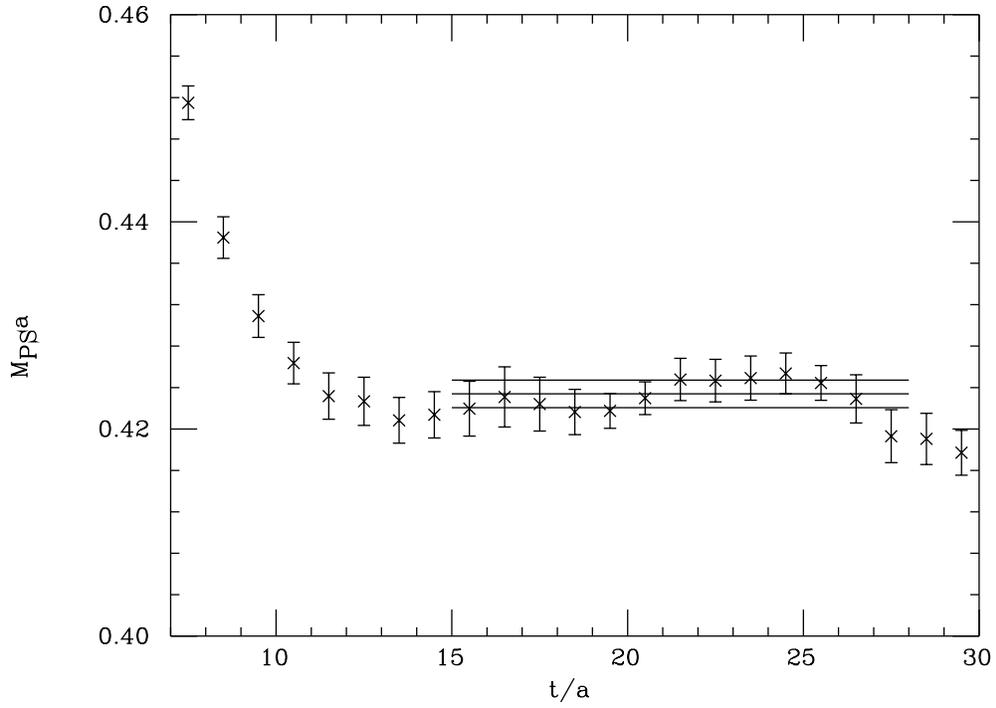}
\caption[]{Effective pseudoscalar mass $M_{PS}$ in lattice units versus time
for lattice W60 and $K=0.1530$.
Dashed lines show the fitted value and its error in the chosen time interval.} 
\protect\label{fig:efm_ps}
\end{figure}

\begin{figure}[htb] 
\putfig{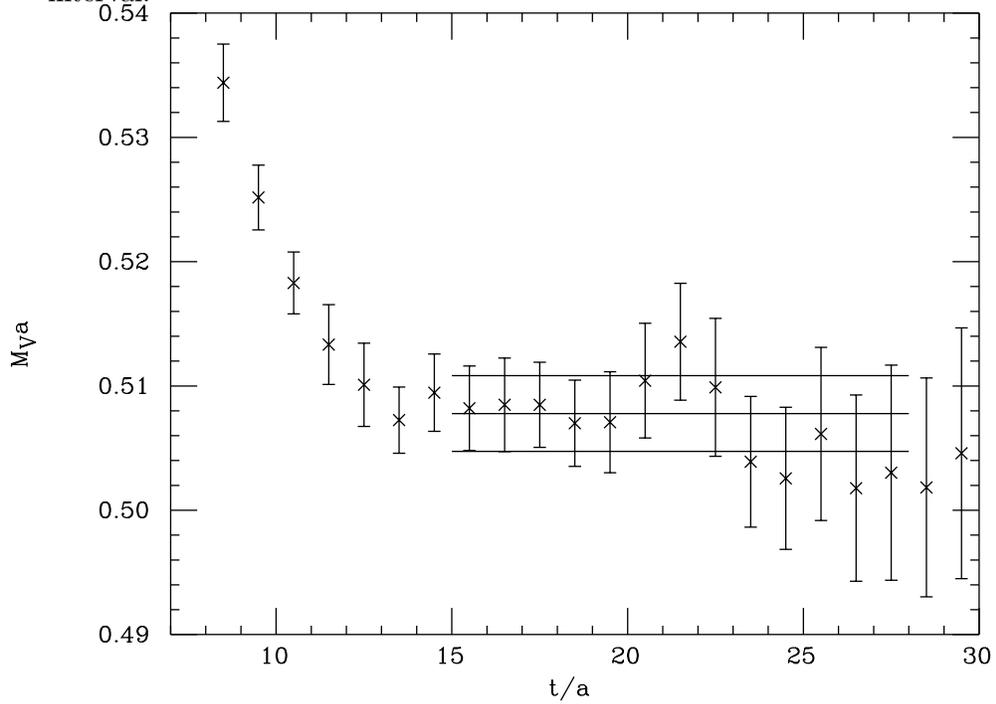}
\caption[]{As in fig.~{\protect\ref{fig:efm_ps}} 
for the vector meson mass $M_V$.}
\protect\label{fig:efm_vi}
\end{figure}

\begin{figure}[htb] 
\putfig{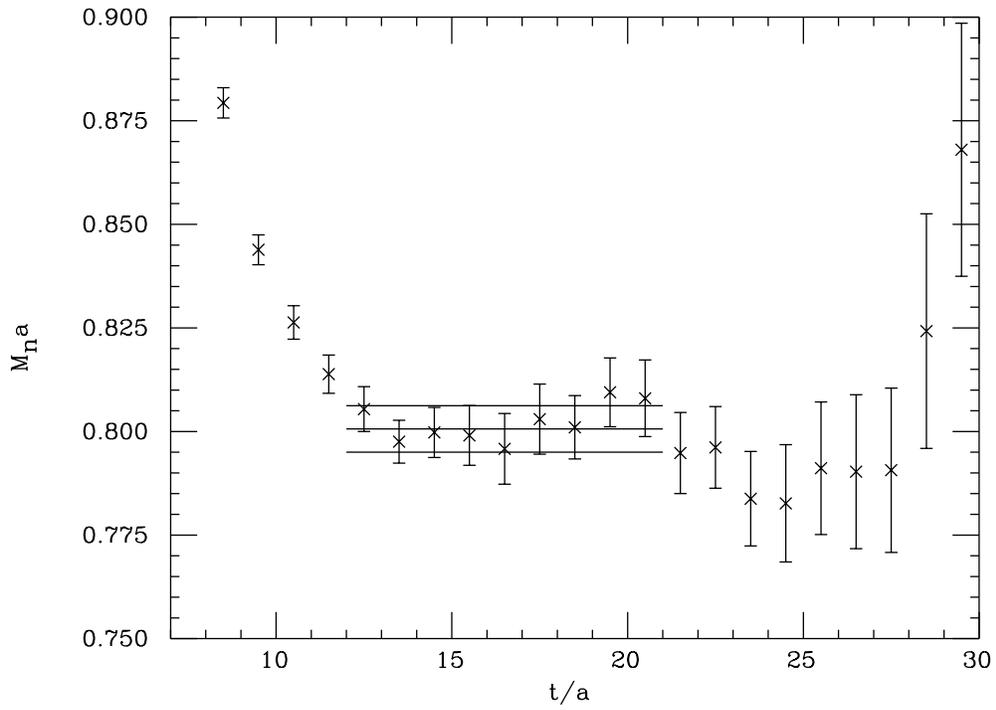}
\caption[]{As in fig.~{\protect\ref{fig:efm_ps}} for the nucleon mass $M_n$.} 
\protect\label{fig:efm_nuc}
\end{figure}

\begin{figure}[htb] 
\putfig{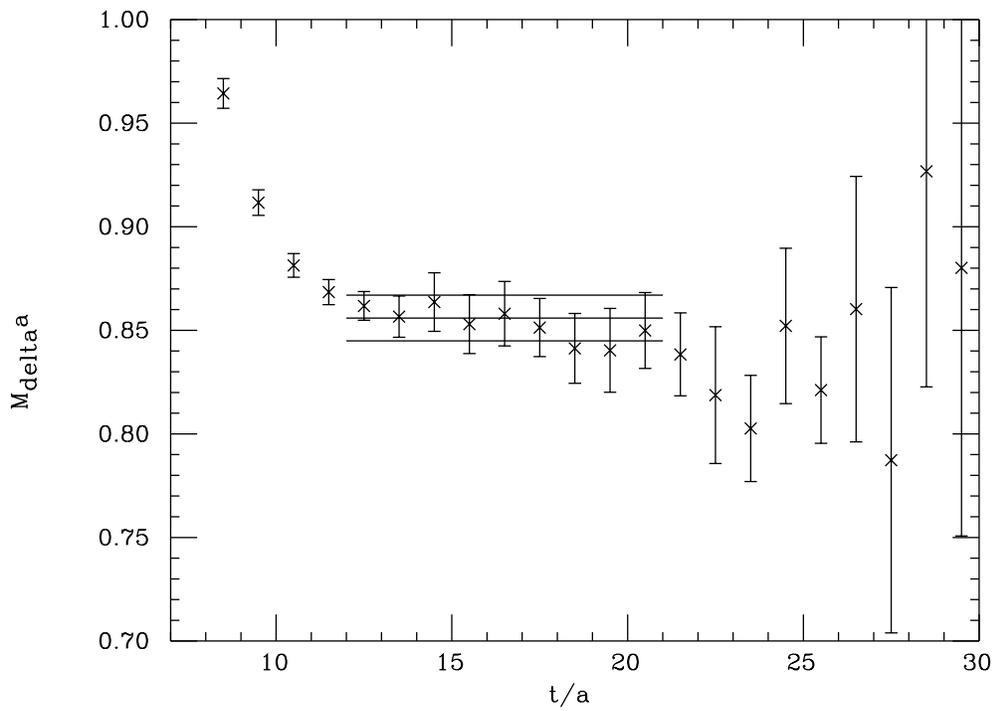}
\caption[]{As in fig.~{\protect\ref{fig:efm_ps}} for delta mass $M_\delta$.}
\protect\label{fig:efm_del}
\end{figure}

Hadron masses and decay constants have been extracted from two-point
correlation functions in the standard way. For the meson masses and decay 
constants we have computed the following propagators:   
\begin{eqnarray}\label{scalare}
G_{55}(t) & = & \sum_{x}\langle P_5(x,t)P_5^\dagger(0,0) \rangle\; ,\nonumber\\
G_{05}(t) & = & \sum_{x}\langle A_0(x,t)P_5^\dagger(0,0) \rangle\; ,
\end{eqnarray}
where
\begin{eqnarray}
P_5(x,t)    & = & i\bar{q}(x,t)\gamma_5q(x,t)\; ,\nonumber\\
A_\mu(x,t)  & = & \bar{q}(x,t)\gamma_\mu\gamma_5q(x,t)\; .\nonumber 
\end{eqnarray}   
and the following propagators of the vector states:
\begin{equation}\label{vettore}
G_{ii}(t) = \sum_{i=1,3}\sum_{x}\langle V_i(x,t)V_i^\dagger(0,0) 
\rangle\; ,
\end{equation}
where 
\[
V_i(x)   =  \bar{q}(x,t)\gamma_i q(x,t)\; .
\]
In order to determine the baryon masses we have evaluated the following 
propagators:
\begin{eqnarray}\label{barioni}
G_n(t) & = & \sum_{x}\langle N(x,t)N^{\dagger}(0,0) \rangle
\; ,\nonumber\\
G_{\delta}(t) & = & \sum_{x}\langle \Delta_\mu(x,t)
\Delta_\mu^{\dagger}(0,0) \rangle\; ,
\end{eqnarray}
where 
\begin{eqnarray}
N              & = &\epsilon_{abc}(u^aC\gamma_5 d^b)u^c\nonumber\\
\Delta_\mu& = & \epsilon_{abc}(u^aC\gamma_\mu u^b)u^c \; .\nonumber
\end{eqnarray}
For some meson correlation functions, non-zero 
momentum values were also studied (see section \ref{par:dispers}).\\
We fit the zero-momentum correlation functions in eqs.~\ref{scalare}, 
\ref{vettore} and \ref{barioni} to a 
single particle propagator with $cosh$ or $sinh$ in the case of mesonic
and axial-pseudoscalar correlation functions
and with an $exp$ function in the case of the baryonic correlation functions
\begin{eqnarray}
G_{55}(t) & = & \frac{Z^{55}}{M_{PS}}\exp(-\frac{1}{2}M_{PS}T)
\cosh(M_{PS}(\frac{T}{2}-t))\; ,\nonumber\\
& & \nonumber\\
G_{ii}(t) & = & \frac{Z^{ii}}{M_V} \exp(-\frac{1}{2}M_{V}T)
\cosh(M_{V}(\frac{T}{2}-t))\; ,\nonumber\\
& & \label{funzfit}\\
G_{05}(t) & = & \frac{Z^{05}}{M_{PS}}\exp(-\frac{1}{2}M_{PS}T)
\sinh(M_{PS}(\frac{T}{2}-t))\; ,\nonumber\\
& & \nonumber\\
G_{n,\delta}(t) & = & C^{n,\delta}
\exp(-M_{n,\delta}t)\; ,\nonumber
\end{eqnarray}
in the time intervals reported in Table~\ref{tab:times}. 
\setlength{\tabcolsep}{.4pc}
\begin{table}       
\label{tab:times}
\caption{Time interval for baryons and mesons used in the fits for all lattices 
and for all momenta.}
\begin{tabular}{ccccccccc}
\hline         
&C60&W60&C62a&W62a&C62b&W62b&W64&C64\\
\hline
\multicolumn{9}{c}{Mesons with zero momentum} \\
$       $ & 15-28 & 15-28 & 18-28 & 18-28 & 18-28 & 18-28 & 24-30 & 24-30 \\
\hline
\multicolumn{9}{c}{Mesons with non-zero momentum} \\
$p^2=1$     & 10-20 & 10-20 & 12-20 & 12-20 & 12-20 & -  & 17-27 & 17-27 \\
$p^2=2$     & 10-20 & 10-20 & 11-16 & 11-16 & 10-15 & -  & 17-22 & 17-22 \\
$p^2=3$     & 10-15 & 10-15 & 11-16 & 11-16 &   -   & -  &   -   &   -   \\
\hline
\multicolumn{9}{c}{Baryons with zero momentum} \\
$       $ & 12-21 & 12-21 & 18-28 & 18-28 & 18-28 & 18-28 & 22-28 & 22-28 \\
\hline
\hline
\end{tabular}
\end{table}
In eqs. \ref{funzfit}, $T$ represents the lattice time extension, the
subscripts $PS$ and $V$ stand for pseudoscalar and vector meson, $n$ and
$\delta$ stand for nucleon- and delta-like baryons\footnote{i.e. $n$ stands for
the nucleon $N$, $\Lambda\Sigma$ or $\Xi$ baryon, and $\delta$ is either
a $\Delta^{++}$ or a $\Omega$.}.
To improve stability, the meson (axial-pseudoscalar)
correlation functions have been symmetrized (anti-symmetrized) around $t=T/2$. 
The time fit intervals are chosen with the following criteria: we fix the lower limit
of the interval as the one at which there is a stabilization of the effective 
mass, 
and, as the upper limit the furthest possible point before the error overwhelms
the signal. We discard the possibility of fitting in a restricted region 
where a plateau is present, as the definition of such a region is
questionable \cite{fukugita}. 
For lattices with highest number of configurations, i.e. W60, W64 and
C64, we confirm that a higher statistics does not lead to a longer or
better (relative to statistical errors) 
plateau \cite{fukugita}. The results of these fits for each run are
reported in Tables \ref{tab:latI}-\ref{tab:latVIII}.
Representative examples of the effective mass plots can be
seen in Figs.~\ref{fig:efm_ps}~-~\ref{fig:efm_del} in the case of W60 
for the heaviest $K = 0.1530$. 
The errors have been estimated by a jacknife procedure, blocking the data in
groups of 10 configurations and we have checked that there are no relevant 
changes in the error estimate by blocking groups of
configurations of different size. We have also attempted to fit the first
excited state mass for the mesons with a two hyperbolic cosine fit but the
results are rather unstable and noisy on all lattices and will not be reported
here. A similar conclusion holds for the baryonic parity partners.

The pseudoscalar and vector decay constants $f_{PS}$ and $1/f_V$ are defined 
through the equations
\begin{eqnarray}
\langle 0|A_0|PS \rangle & = & i \frac{f_{PS}}{Z_A} M_{PS}\; ,\\
\langle 0|V_i|V,r \rangle & = & \epsilon^r_i \frac{M_V^2}{f_V Z_V}\; ,
\end{eqnarray}
where $\epsilon^r_i$ is the vector-meson polarization, $M_{PS}$ and $M_V$ are
the pseudoscalar and vector masses and $Z_{V,A}$ are the renormalization
constants.
We extract $f_{PS}$ from the ratio
\begin{eqnarray}
R_{f_{PS}}(t) & = & Z_A \frac{G_{05}(t)}{G_{55}(t)}\nonumber\\
& \longrightarrow & Z_A \frac{Z^{05}}{Z^{55}}\tanh(M_{PS}(\frac{T}{2}-t))\nonumber\\  
& = & Z_A \frac{\langle 0|A_0|P \rangle}{\langle 0|P_5|P\rangle}
      \tanh(M_{PS}(\frac{T}{2}-t))\nonumber\\
& = & \frac{f_{PS} M_{PS}}{\sqrt{Z^{55}}}\tanh(M_{PS}(\frac{T}{2}-t))\; ,
\end{eqnarray}
and the vector-meson decay constant is obtained directly from the parameters of
the fit to $G_{ii}(t)$, eqs. \ref{funzfit}:
\begin{eqnarray}
\frac{1}{Z_V f_V} = \frac{\sqrt{Z^{ii}}}{M_V^2} \; .
\end{eqnarray}
The results of these fits for each run are reported in Tables
\ref{tab:latI}~-~\ref{tab:latVIII}.


\section{Extraction of physical quantities}
\label{new_method}
Once the hadronic correlation functions have been fitted, and the lattice
masses and matrix elements extracted, the `conventional method' of determining
physical quantities relies on a number of chiral extrapolations. The first
extrapolation is in the lattice pseudoscalar mass squared, $(M_{PS} a)^2$,
against $1/K$. This sets the critical kappa value $K_c$ corresponding to zero
quark mass. A second chiral extrapolation is performed of $(M_V a)$ to zero
quark mass. Using the experimental value of the $\rho$ mass, $M_V(K_c) =
770~MeV$, the lattice spacing $a$ is determined (assuming a massless pion).

\begin{figure}[htb] 
\putfig{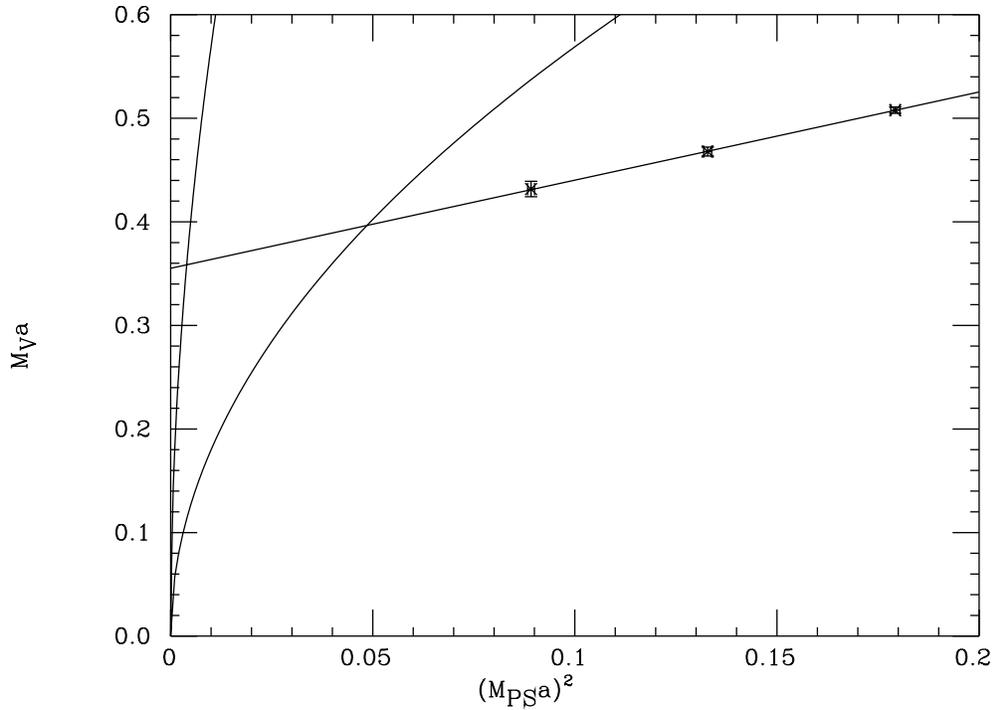}
\caption[]{Values of $M_V$ versus $M_{PS}^2$ in lattice units and the fitting
line for W60. 
The curves represent the equations $M_V a = C \sqrt{(M_{PS} a)^2}$ with
$C=C_{ll}$ (leftmost) and $C=C_{sl}$ (rightmost) defined in section
\ref{new_method}.}
\protect\label{fig:mv_v_mp^2}
\end{figure}

\begin{figure}[htb] 
\putfig{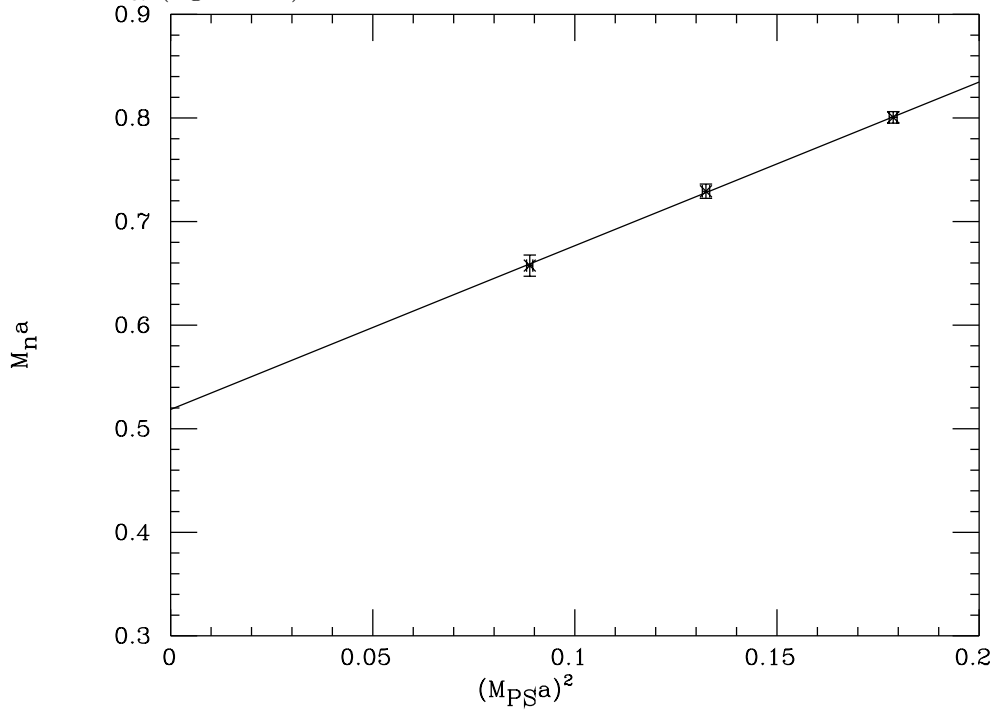}
\caption[]{Values of $M_n$ versus $M_{PS}^2$ in 
lattice units and the fitting line for W60.}
\protect\label{fig:mn_v_mp^2}
\end{figure}

\begin{figure}[htb] 
\putfig{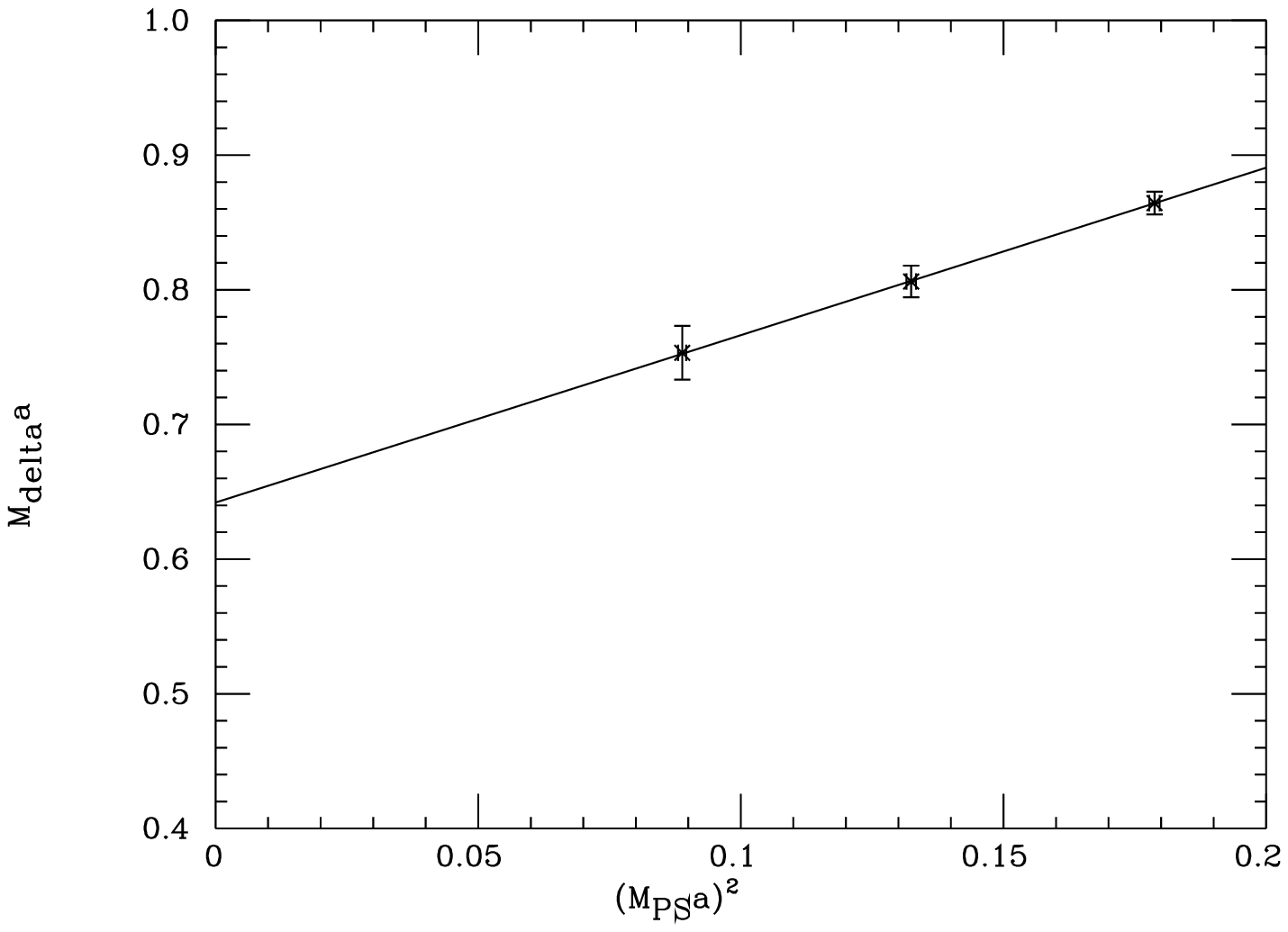}
\caption[]{As in fig.~{\protect\ref{fig:mn_v_mp^2}} for $M_\delta$.}
\protect\label{fig:md_v_mp^2}
\end{figure}

\begin{figure}[htb] 
\putfig{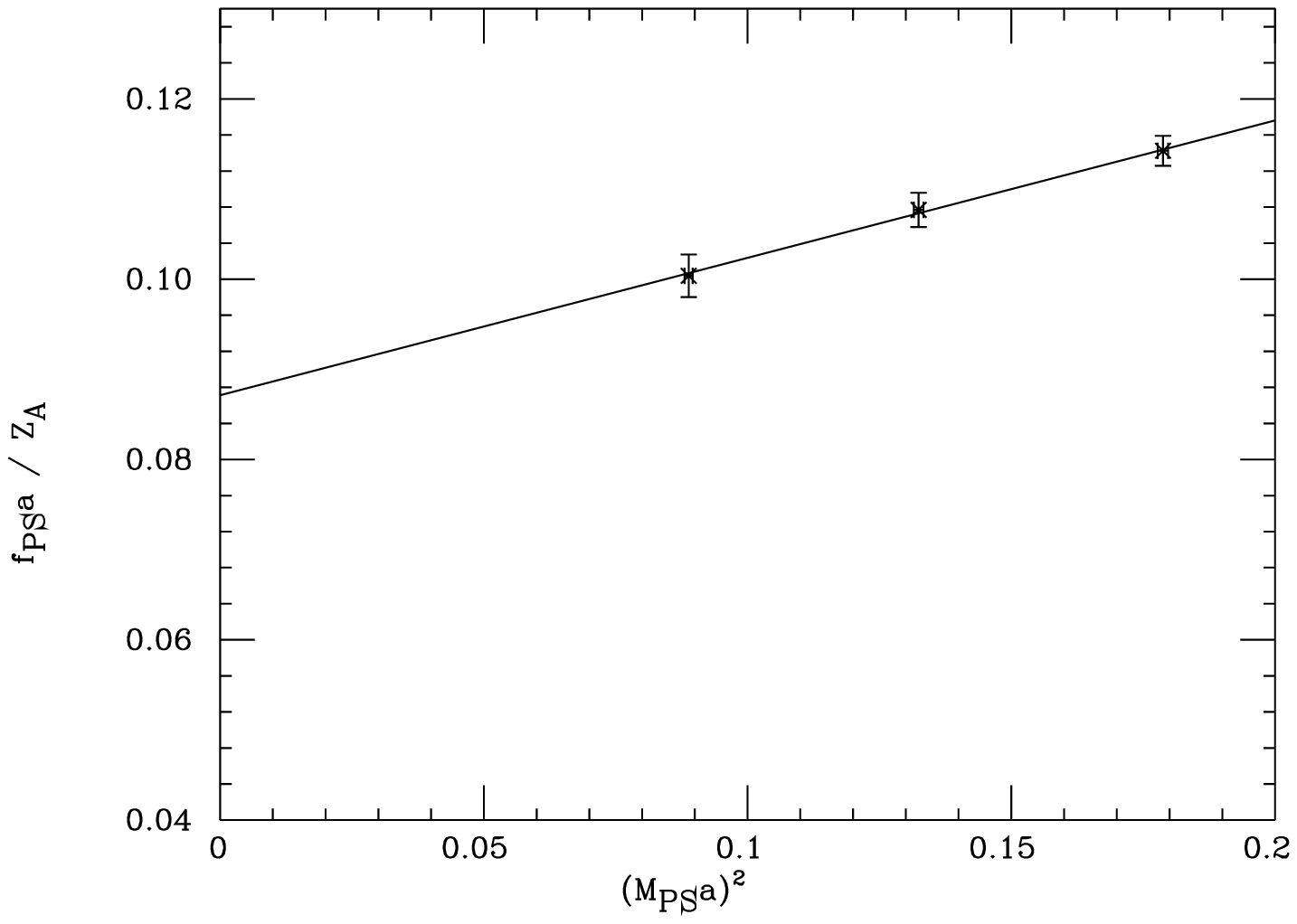}
\caption[]{As in fig.~{\protect\ref{fig:mn_v_mp^2}} for $\displaystyle\frac{f_{PS}}{Z_A}$.}
\protect\label{fig:fp_v_mp^2}
\end{figure}

\begin{figure}[htb] 
\putfig{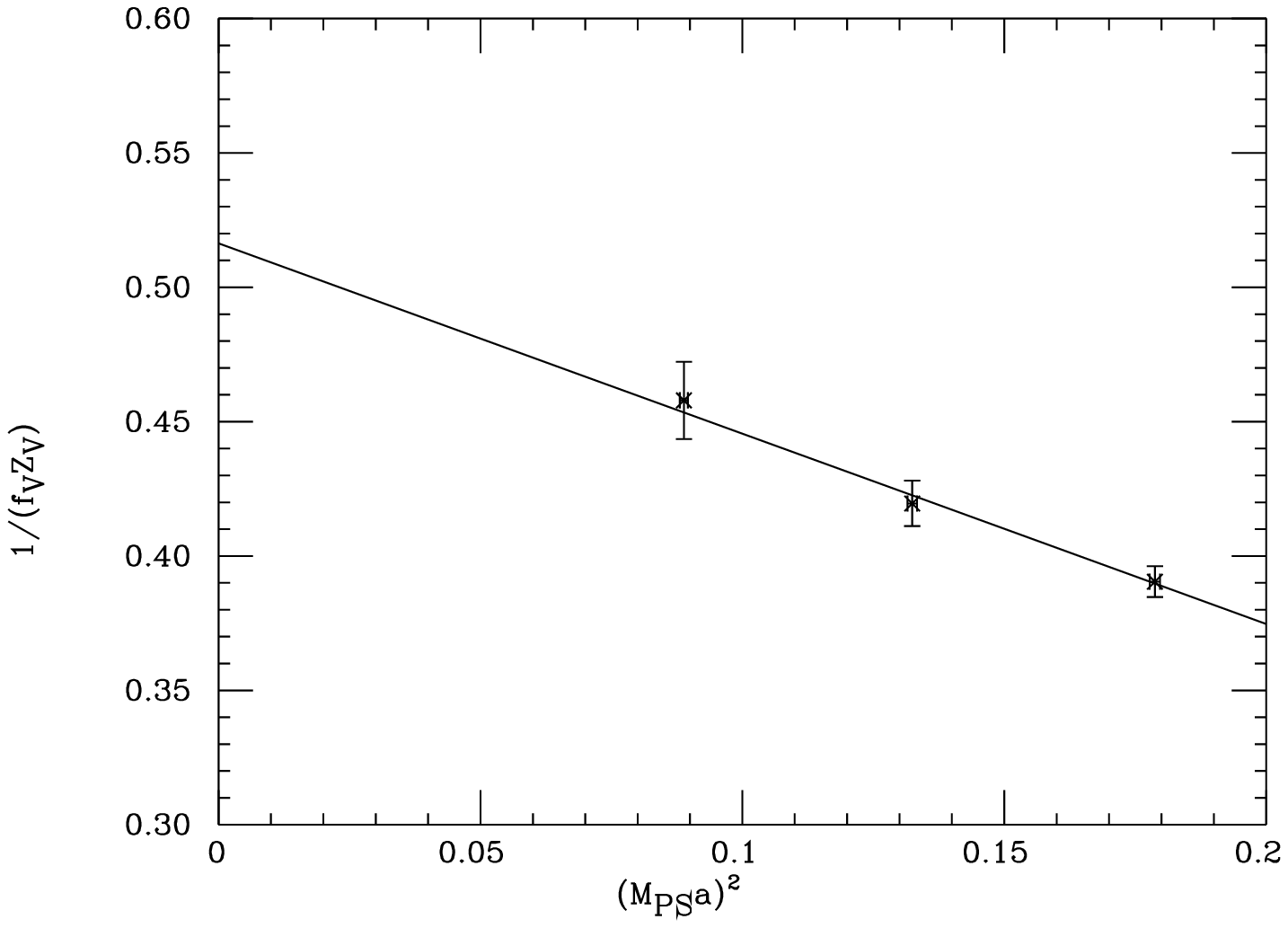}
\caption[]{As in fig.~{\protect\ref{fig:mn_v_mp^2}} for $\displaystyle\frac{1}{f_V Z_V}$.}
\protect\label{fig:fv_v_mp^2}
\end{figure}

This conventional method relies on an 
extrapolation from a region where data has
been obtained to a region well beyond the reach of direct simulations, 
assuming that the pseudoscalar mass follows the
predictions of PCAC, and that other physical quantities scale linearly
in the quark mass\footnote{It is unlikely that the problem of
`chiral logs' in the quenched approximation be visible at the values 
of the simulated quark masses \cite{sharpe}.}.
This problem suggests one should extract as much physics as  
possible from the  `strange' region and to use chiral extrapolations
only where absolutely necessary. 
The method we use to achieve this aim is outlined below.
\begin{itemize}
\item We define the following five `lattice physical planes' for masses
and decay constants: \hskip 5mm ($M_V a$, $(M_{PS} a)^2$),  ($f_{PS}
a/Z_A$, $(M_{PS} a)^2$), ($1/(f_{V} Z_V)$, $(M_{PS} a)^2$), 
($M_n a$, $(M_{PS} a)^2$) and ($M_\delta a$, $(M_{PS} a)^2$). We plot the
lattice data for each $K$ value used in the simulation on these planes . (See
Figs.~\ref{fig:mv_v_mp^2}~-~\ref{fig:fv_v_mp^2}.)
\item On the $(M_V a\; ,(M_{PS}a)^2)$ plane we impose the physical ratio 
$C_{sl} = M_{K^*}/M_{K}$.
This corresponds to find the intercept of the curve
\begin{equation}
M_V a = C_{sl} \sqrt{(M_{PS} a)^2},
\label{eq:C}
\end{equation}
with the linear fit of the lattice data (i.e. the right-hand
curve in Fig.~\ref{fig:mv_v_mp^2}). Lacock and Michael \cite{michael} have
used this approach to define the quantity $J$. (See also \cite{mm} and
\cite{weingarten}.) The intercept of this curve and the simulated data
defines $M_K a$ and $M_{K^*} a$ but only one of these is an
independent prediction.
\item We now use the value of $M_K a$ determined above to read off the
lattice meson decay constants, $(f_K a/Z_A)$ and $(f_{K^*} Z_V)^{-1}$
from the corresponding $f_{PS}$ and $f_V$ planes. At this stage we have
the following 3 independent predictions: $(M_{K^*} a)$, $(f_K a/Z_A)$
and $(f_{K^*} Z_V)^{-1}$. These were all obtained from the quark mass
region actually simulated (see Figs~\ref{fig:mv_v_mp^2},
\ref{fig:fp_v_mp^2} and \ref{fig:fv_v_mp^2}), and hence they do not rely
on a chiral extrapolation.\footnote{Depending on the choice of  $K$,
a modest linear extrapolation in $M_{PS}^2$ may be required. 
The range of this extrapolation is small and so should not cause problems.}
\item The lattice spacing $a$, can simply be set by comparing either
$(M_{K^*} a)$ or $(f_K a/Z_A)$ with their experimental value.
($f_{K^*}$ cannot be used since it is dimensionless.)
Of these two quantities, we choose $M_{K^*}$ to set the
scale whenever we need to quote results in physical units. This is
because $M_{K^*}$ is determined experimentally more accurately than
$f_K$, and it has no ambiguities associated with a renormalisation
constant. However, for completeness, we list in Table~\ref{tab:inva}
the $a^{-1}$ values obtained from all the quantities.
\end{itemize}

In this analysis for $M_{K^*} a$, $f_K a / Z_A$ and $(f_{K^*} Z_V)^{-1}$ we have
assumed that the quadratic $SU(3)-$flavour breaking effects in the meson masses
or decay constants can be safely neglected in the range of masses explored 
here. This means that we assume that the masses and decay constants
for mesons, with both degenerate, and non-degenerate constituent quarks, lie on
the same universal curve. Studies have shown this to be an entirely reasonable
hypothesis for quarks around the strange mass \cite{ukqcd93}.

To proceed to obtain other physical quantities, we need the additional
input of quark model considerations, and chiral extrapolations.
Therefore the determination of these additional quantities suffers from
the same criticisms as outlined above for the conventional approach.
Consider the hadronic physical quantity $M$ (where $M = M_V$,
$f_{PS}$, $f_V, M_n$ or $M_\delta$). As a first approximation, the
dependence of $M$ on the hadron's quark masses can be
approximated by
\begin{equation}
M a = A_{M} + B_{M} \sum_{q=1}^{Q} m_q a,
\label{eq:qm}
\end{equation}
where the $m_q$ are the masses of the constituent quarks, and $Q=2,3$
for mesons, baryons. This is the usual assumption of the quark model.
In order to continue we need the additional assumption of PCAC
\begin{equation}
(M_{PS}a)^2 = D (m_1a + m_2a).
\label{eq:pcac}
\end{equation}
Combining eqs. \ref{eq:qm} and \ref{eq:pcac} and eliminating the quark
masses we have in the case of $M = M_V$,
\begin{eqnarray}
M_V a &=& A_{M_V} + \frac{B_{M_V}}{D} (M_{PS} a)^2.
\label{eq:mv}
\end{eqnarray}
Assuming this remains valid from the simulation region down to (near-)
zero quark masses, we can obtain a prediction for $M_\rho a$ and
$(M_\pi a)^2$ as follows. 
\begin{itemize}
\item On the $(M_V a\; ,(M_{PS}a)^2)$ plane we impose the physical ratio 
$C_{ll} = M_{\rho}/M_{\pi}$. The intercept of this curve and the linear 
extrapolation of the simulated data
defines $M_\rho a$ and $M_{\pi} a$ (only one independent).
\item We now use the value of $M_\pi a$ determined above to read off the
lattice meson decay constants, $(f_\pi a/Z_A)$ and $(f_{\rho} Z_V)^{-1}$
from the corresponding $f_{PS}$ and $f_V$ planes. 
\end{itemize}
From Eq.(\ref{eq:pcac}) we immediately have
\begin{equation}
(M_{\eta_{ss}} a)^2 = 2 (M_K a)^2 - (M_\pi a)^2,
\end{equation}
where $\eta_{ss}$ is the hypothetical pseudoscalar 
meson with quark content
$s\overline{s}$ assuming no mixing with other states, for whose
`experimental' mass we have $M_{\eta_{ss}}~=~0.686~GeV$. 
Using the lattice value of $M_{\eta_{ss}} a$, we can determine $M_\phi$ and
$(f_\phi Z_V)^{-1}$ from Figs.~\ref{fig:mv_v_mp^2} \& \ref{fig:fv_v_mp^2}.
As a check of the assumption of PCAC and Eq.~\ref{eq:qm}, we plot in
Fig.~\ref{fig:mv_mp2} the lattice values of $M_V/M_{K^*}$ against
$M_{PS}^2/M_{K}^2$ together with their experimental values.
The data lie on a straight line to remarkable accuracy. 

For baryonic masses, the situation is more complicated.
Eq.(\ref{eq:qm}) can be written as:
\begin{eqnarray} \label{eq:mn}
M_n a &=& A_{M_n} + B_{M_n} ( m_1 a + m_2 a + m_3 a ), \;\;\;\; \\ \nonumber
M_\delta a&=& A_{M_\delta} + B_{M_\delta} ( m_1 a + m_2 a + m_3 a ).
\end{eqnarray}
Combining these with eq.~(\ref{eq:pcac}) leads to
\begin{eqnarray}
M_N a      &=& A_{M_n} + \frac{B_{M_n}}{D}
 \left[ \frac{3}{2} (M_{\pi} a)^2        \right] \nonumber\\
M_{\Lambda\Sigma} a &=& A_{M_n} + \frac{B_{M_n}}{D}
 \left[ (M_K a)^2 + \frac{1}{2} (M_\pi^2 a)^2  \right] \nonumber\\
M_\Xi a    &=& A_{M_n} + \frac{B_{M_n}}{D}
 \left[ 2 (M_K a)^2 - \frac{1}{2} (M_\pi a)^2\right] \nonumber\\
M_\Delta a &=& A_{M_\delta} + \frac{B_{M_\delta}}{D}
 \left[ \frac{3}{2} (M_\pi a)^2 \right] \nonumber\\
M_\Omega  a&=& A_{M_\delta} + \frac{B_{M_\delta}}{D}
 \left[ 3 (M_K a)^2 - \frac{3}{2} (M_\pi a)^2 \right].\nonumber
\end{eqnarray}

where

\be
M_{\Lambda\Sigma} = \frac{1}{4} M_\Sigma + \frac{3}{4} M_\Lambda
\ee

The baryonic masses can be extracted by fitting the data for $M_n a$ and
$M_\delta a$ against $(M_{PS} a)^2$ to a straight line and using the values 
of $M_\pi a$ and $M_K a$ already determined, (see
Figs.~\ref{fig:mn_v_mp^2} and \ref{fig:md_v_mp^2}). Due to the limited number
of $K$ values, we do not attempt the fit to more complicated functions \cite{chiral}. We have
also checked that a quadratic fit does not give better $\chi^{2}$. 
Our predictions of the initial 3 quantities which do not require any
chiral extrapolation have then been extended to include the following quantities
which do require chiral extrapolations and quark model considerations:
\[
\!\!\!\!
M_\rho, M_\phi, f_\pi / Z_A, (f_\rho Z_V)^{-1},  (f_\phi Z_V)^{-1}, 
M_N, M_{\Lambda\Sigma}, M_\Xi, M_\Delta \mbox{and} M_\Omega.
\]
The method outlined above has the following properties:
\begin{itemize}
\item We use only the lattice physical planes and not the unphysical 
planes where the $x-\/$axis is the quark mass $1/K$. These unphysical
planes will only be required when we need to determine unphysical
quantities such as the quark masses.
\item This method allows us to fix the value of $M_{K^*} a$, and therefore
$a^{-1}$ without the need of any chiral extrapolation. It also enables
us to determine $f_K$ and $f_{K^*}$ without any extrapolation.
\item With the additional assumption of PCAC (Eq.(\ref{eq:pcac})) and
the quark model (Eq.(\ref{eq:qm})) we can derive the rest of the meson
spectrum and decay constants, and the baryon spectrum.
\item The conventional approach can be seen as the limit
$C_{ll}\rightarrow\infty$ of this approach.
\end{itemize}
In the next section we use the above method to determine the physical
quantities from the lattice simulations.

\section{Results}\label{results}
This section briefly lists the results obtained using the method described in
the previous section. Tables~\ref{tab:latI}~-~\ref{tab:latVIII} contain the
lattice data at each $K$ value for $M_{PS} a$, $M_{V} a$, $M_n a$, $M_\delta
a$, $f_{PS} a/Z_A$ and $(f_V Z_V)^{-1}$ from the fits outlined is Section
\ref{lat_details}. Also listed are the $K_c$ values
(obtained using Eq.(\ref{eq:pcac}) with $m_q=1/2 (1/K_q - 1/K_c)$ and the
chiral extrapolants using a linear fit with $M_{PS}^2$. All the mesonic and
baryonic masses are also plotted in figs~\ref{fig:mv_mp2}-\ref{fig:mdelta_mp2}
and show a very good consistency, within the errors, among the various
lattices. This shows that we observe a scaling behaviour in our data.

\subsection{Hadron masses and decay constants}
Using the method described in Sect.~\ref{new_method}, the meson masses
and decay constants  (in lattice units) are displayed in
Table~\ref{tab:mesons} and \ref{tab:decay}. The inverse lattice
spacing, $a^{-1}$, obtained from the
experimental values is displayed in Table~\ref{tab:inva}.\\
Similarly the baryon masses, are displayed in Table~\ref{tab:baryons}. Again,
for comparison, the $a^{-1}$ values obtained using these masses are
shown in Table~\ref{tab:inva}.
Note that all errors quoted in Tables~\ref{tab:mesons}, \ref{tab:decay} and
\ref{tab:baryons} are statistical only.
In Tables~\ref{tab:mesonsGeV}~-~\ref{tab:decaysGeV}, a listing is given of 
the lattice predictions
of the mass spectrum and decay constants from C60~-~C64. The scale
was set from the $K^*$ mass.

Comparing the clover meson masses at $\beta=6.2$ from lattices
C62a, C62b and the
UKQCD results  \cite{ukqcd93} in fig.~\ref{fig:vol_62}, we find very good
agreement with UKQCD for Lattice C62a. For C62b, while we find unexpectedly
larger fluctuations in the comparison, we seem to get a somewhat lower mass;
there may be some residual finite size effects in the vector meson mass, even
though one would rather expect a higher mass in this case. This problem may
also be present for the $\beta=6.4$ data for which the physical volume is the
same as in C62b. Further investigation at larger  lattice sizes is necessary
to make the situation clearer.
At $\beta=6.0$ we can also compare our Wilson data with \cite{parisi},
\cite{fukugita} and \cite{gupta}, see 
Table~\ref{tab:latII} and fig.~\ref{fig:vol_60}. The agreement 
is very good for the pseudoscalar data for which we have $K$ values in 
common. Our data for the 
vector mass at the lightest quark mass is slightly higher.
We do not have a clear explanation for this. We only note
that if we had fitted our data in the $t$ region in the 
interval $11~-~21$ as done in fig.~2  of ref.~\cite{gupta}
our value for the vector mass would have been 
$M_V=0.427(4)$ which is in much better agreement with their
value. 
On the other hand, for the baryon data, we find very good 
agreement with the old APE \cite{parisi} data, while we find
slightly higher values and larger fluctuations 
when comparing with JLQCD \cite{fukugita} and 
LANL \cite{gupta} data as shown in fig.~\ref{fig:baryword}.

As can be seen from Tables~\ref{tab:mesonsGeV}~-~\ref{tab:baryonsGeV}, there is
good consistency of the physical predictions between the different simulations
to within statistical errors. Considering the fact that the errors shown in
Tables~\ref{tab:mesonsGeV}~-~\ref{tab:baryonsGeV} are  statistical only, and
that these can often be underestimated \cite{fukugita}, the agreement is
remarkable.

To compare the lattice decay constants with the experimental ones, we have used
a `boosted' one-loop form of the renormalisation constants
\cite{zwang}~-~\cite{gabrielli2}:
\begin{eqnarray} 
\mbox{Wilson action}\; & Z_A & =  1 - 0.134 g_{\overline{MS}}^2  \nonumber\\ 
                       & Z_V & =  1 - 0.174 g_{\overline{MS}}^ 2  \nonumber\\ 
\mbox{Clover action}\; & Z_A & =  1 - 0.0177 g_{\overline{MS}}^2 \nonumber\\ 
                       & Z_V & =  1 - 0.10  g_{\overline{MS}}^2  \nonumber
\end{eqnarray}
where $g_{\overline{MS}}^2 = 6/\beta_{\overline{MS}}$,
with \cite{elkhadra}~-~\cite{lepage}
\begin{equation}
\beta_{\overline{MS}} = <U_{plaq}> \beta + 0.15.
\end{equation}

The results are shown in Table~\ref{tab:decaysGeV} and
figs.~\ref{fig:fpsmv_mp2wil} \& \ref{fig:fvmp2clo}.
For the ratio $f_{PS}/{M_V}$ in
figs.~\ref{fig:fpsmv_mp2wil}-figs.~\ref{fig:fpsmv_mp2clo}, we see again a
global consistency  of our different data both for Wilson and the SW-Clover
action. The UKQCD data \cite{ukqcd93}, also reported in
fig.\ref{fig:fpsmv_mp2clo},  with the same renormalization decay constant $Z_A$
disagree with ours for two of the three points where the comparison is
possible. For the third point there is agreement but within much larger error
bars. We do not see a strong $a$ dependence for either actions comparing
lattices at different $\beta$ (C60, C62a and W60, W62a). This confirms
statements made in Sect.~\ref{lat_details} about the statistical presence of
scaling in our data. The experimental points do lie quite well in the band
extrapolated/interpolated from the lattice data.

The values of the pseudoscalar decay constants in Table~\ref{tab:decaysGeV}
show larger deviations from the experimental data than the ratio
$f_{PS}/{M_V}$. It is clear that a cancellation of systematic error occurs in
this ratio. Still there is an agreement within 1.5 standard deviations with the
experimental data apart from the W60 value which is quite high. In particular
it is higher than the value obtained for the Wilson action at $\beta = 6$ in
ref.~\cite{guptafpai}. There the authors use different renormalization schemes
than the one used here but the discrepancy is too large to be attributed to
this.

The situation is more delicate for the vector decay constant for which a
dependence on the volume and $a^{-1}$ may be present but with our data it 
would be difficult to disentangle the two effects. In particular in 
figs.~\ref{fig:fvmp2wil}, \ref{fig:fvmp2clo} and Table~\ref{tab:decaysGeV} the
values for C62a  and C62b (SW-Clover on different physical volumes), C62a and  W62a
(SW-Clover and Wilson with the same physical  volume), W60 and W62a ( Wilson with the
same physical volume but  different values of $\beta$) and C60 and C62a
(SW-Clover on the  same volume and different $\beta$) seem to show this problem. The
extrapolated values for $1/f_V$ suffer very much due to these effects. The
`strange' vector decay constants seem to be slightly more stable also because
no extrapolation is needed.
\subsection{Meson Dispersion relations}\label{par:dispers}
For all the lattices except for W62b the 
non-zero momentum meson correlation functions were also computed. We can study
discretization errors and finite volume effects testing which dispersion
relation is best verified from data. We have tested the following dispersion
relations
\ba
E^2 & = & p^2 + M^2  \; ,\label{disA}\\
\sinh^2(\frac{E}{4}) & = & \sin^2(\frac{p}{4}) + \sinh^2(\frac{M}{4})\; ,\\
\sinh^2(\frac{E}{2}) & = & \sin^2(\frac{p}{2}) + \sinh^2(\frac{M}{2})\; ,\label{disc}\\ 
\sinh^2(E) & = & \sin^2(p) + \sinh^2 M\; ,
\ea    
where (\ref{disA}) is the continuum relation, while the others are lattice
dispersion relations from different choices of lattice action.
$M=E(\vec{p}=0)$ is the mass. 
From the lattice we have extracted the energy 
$E(\vec{p})$ of the
states fitting the non-zero momentum correlation functions with a $cosh$
function. The results we obtained for $E(\vec{p})$ for pseudoscalar and vector 
mesons are reported in Table~\ref{tab:disp}. 
We see only small differences among the various
dispersion relations for the quark masses we have simulated, but eq.~(\ref{disc})
seems to be favoured \cite{serone}. This is consistent with the conclusion of 
ref.~\cite{gupta} which refers however to much larger quark masses.    
\subsection{Strange Quark Mass}
Lattice QCD is in principle able to predict the mass of any quark by fixing to
its experimental value the mass of a hadron containing a quark with the same
flavour. The `bare' lattice quark mass $m(a)$ can be extracted directly from
lattice simulations and can be related to the continuum mass 
$m^{\overline{MS}}(\mu)$ renormalized in the minimal-subtraction dimensional
scheme through a well-defined perturbative procedure
\cite{qmass2}~-~\cite{bochicchio} .  Following ref.~\cite{qmass1}
\[
m^{\overline{MS}}(\mu) = Z_m^{\overline{MS}}(\mu a)m(a)
\] 
where $m(a)$ is the bare lattice quark mass related to the hopping parameter as
\be\label{ultima}
m(a) a = \frac{1}{2}(\frac{1}{K}~-~\frac{1}{K_c})\; . 
\ee 
and $Z_m^{\overline{MS}}(\mu a)$ is the mass renormalization constant at scale
$\mu$ which we choose to be $2~GeV$. The hopping parameters and the strange
quark masses reported in Tables~\ref{tab:kappas} and \ref{tab:quarkm} are
extracted using the eq. (\ref{ultima}) and (\ref{eq:pcac}) from the lattice
values $M_\pi a$ and  $M_K a$.
The error on $m^{\overline{MS}}(\mu)$ is estimated as in ref.~\cite{qmass1}
taking into account the spread due to different definitions of the strong
coupling constant. This leads to somewhat larger errors for the Wilson data than
for the Clover ones due to the smaller fluctuation of the different
renormalization constants in the improved case.
There is a good consistency among the values coming from the different
lattices. Fig.~\ref{fig:qmas} shows that we do not see any dependence on $a$
within the errors in the Wilson data and a mild tendency in the clover data to
decrease with increasing $\beta$.
We therefore conclude that any $O(a)$ effects present are beneath the
level of statistics, and/or hidden among finite volume effects.
We then extract the average value of the
strange quark mass without any extrapolation and get:
\[
m_s^{\overline{MS}}(\mu = 2 GeV) = 122 \pm 20 MeV
\]
which is in agreement with the result of ref.~\cite{qmass1}. It is also
compatible with the value $m_s^{\overline{MS}}(\mu = 2 GeV) = 90 \pm 20 \;
MeV$ of
ref.~\cite{guptamstrange}, but one should take into account that this value
comes from an analysis on various Wilson and Staggered lattices at different
values of $\beta$ and an extrapolation in $a$.
The quark masses can also be extracted from the Ward identity for the axial
vector current. A full analysis of the light quarks masses from both methods
will be presented in a forthcoming paper.
\subsection{J value(s)}
\begin{figure}[htb] 
\putfig{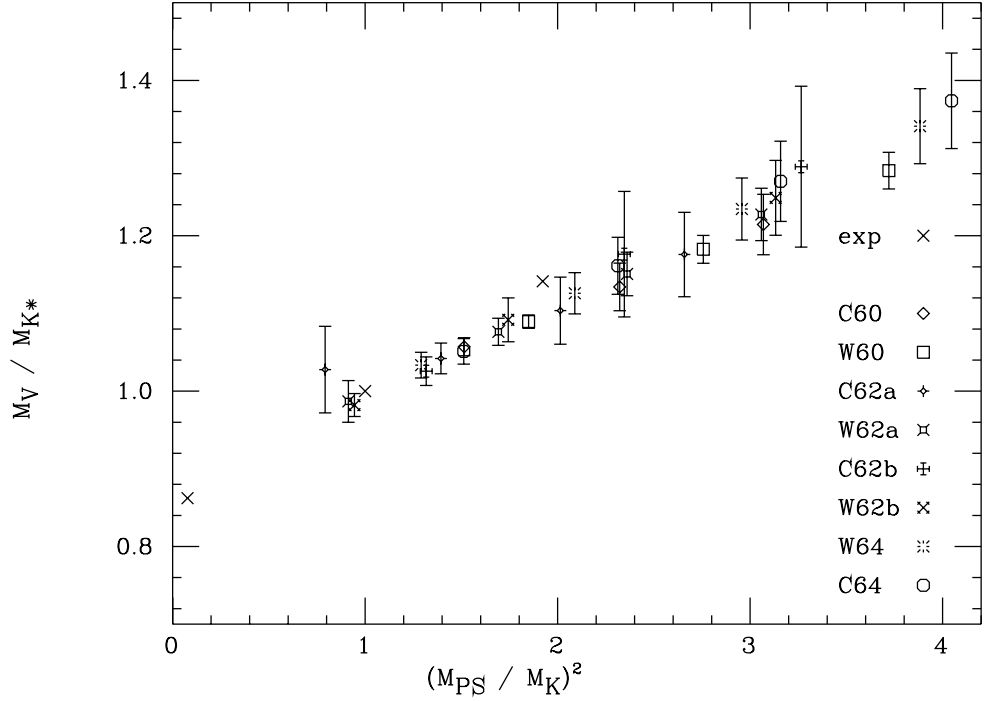}
\caption[]{Mass of vector meson $M_V$ versus the squared pseudoscalar mass
$M_{PS}^2$  for all lattices. $M_V$ and $M_{PS}$ normalized to the lattice
value of $M_{K^*}$ and  $M_{K}$ respectively.}
\protect\label{fig:mv_mp2}
\end{figure}

\begin{figure}[htb] 
\putfig{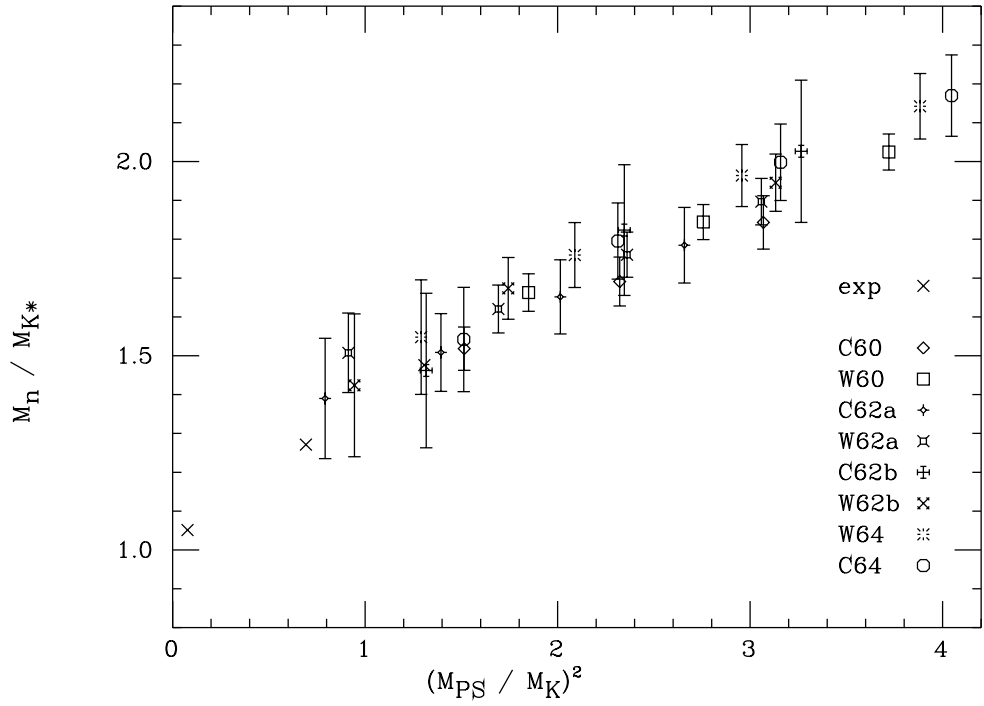}
\caption[]{As in fig.~{\protect\ref{fig:mv_mp2}} for $M_n$.}
\protect\label{fig:mb_mp2}
\end{figure}

\begin{figure}[htb] 
\putfig{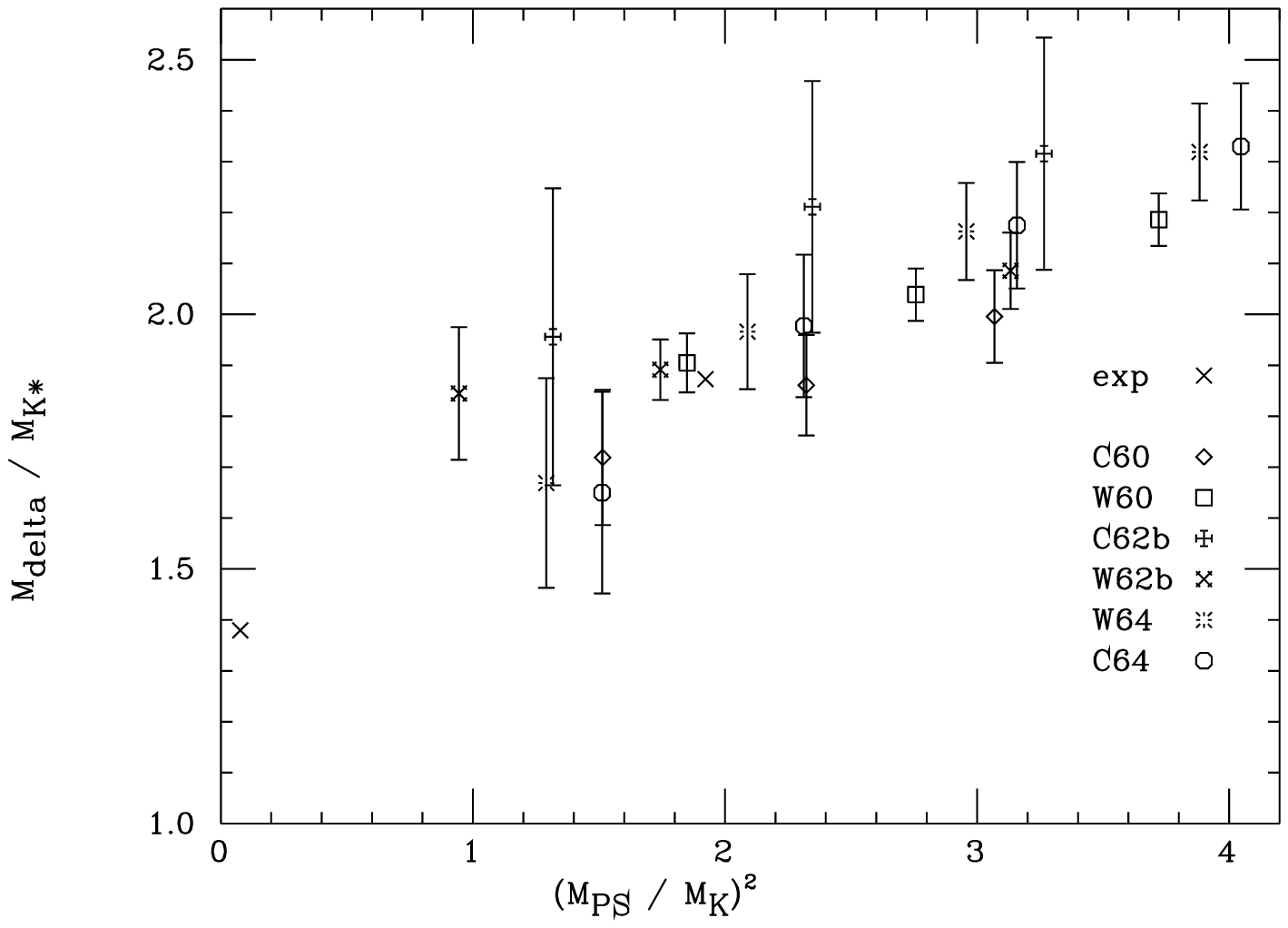}
\caption[]{As in fig.~{\protect\ref{fig:mv_mp2}} for $M_\delta$.}
\protect\label{fig:mdelta_mp2}
\end{figure}

\begin{figure}[htb] 
\putfig{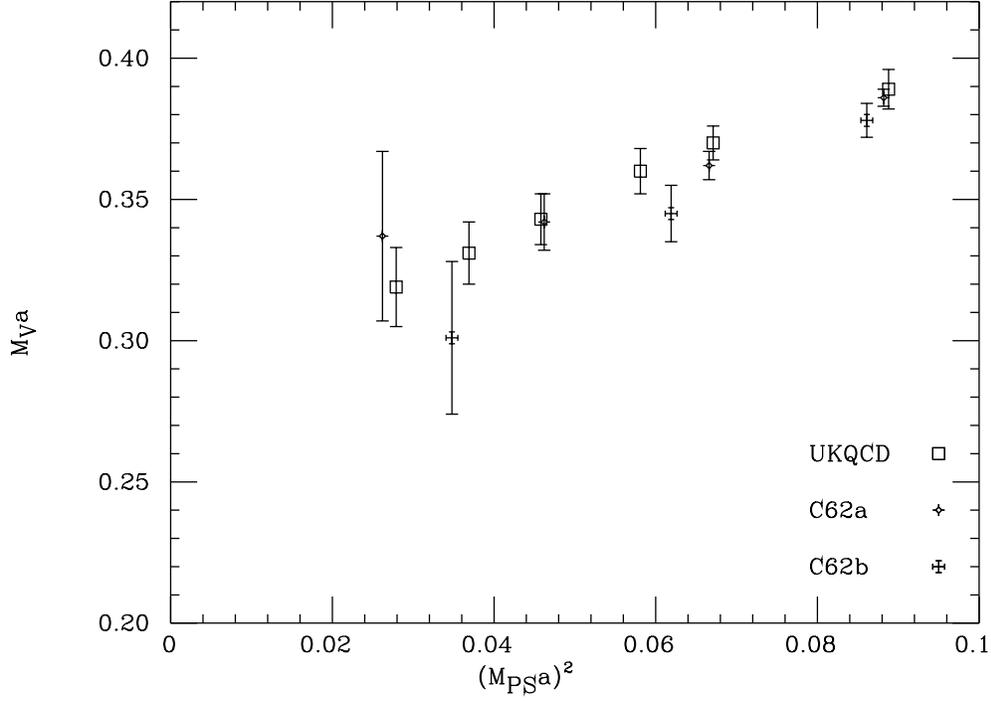}
\caption[]{Vector meson masses in lattice units at $\beta=6.2$ for SW-Clover action
from UKQCD \cite{ukqcd93}, Lattice C62a and Lattice C62b.}
\protect\label{fig:vol_62}
\end{figure}

\begin{figure}[htb] 
\putfig{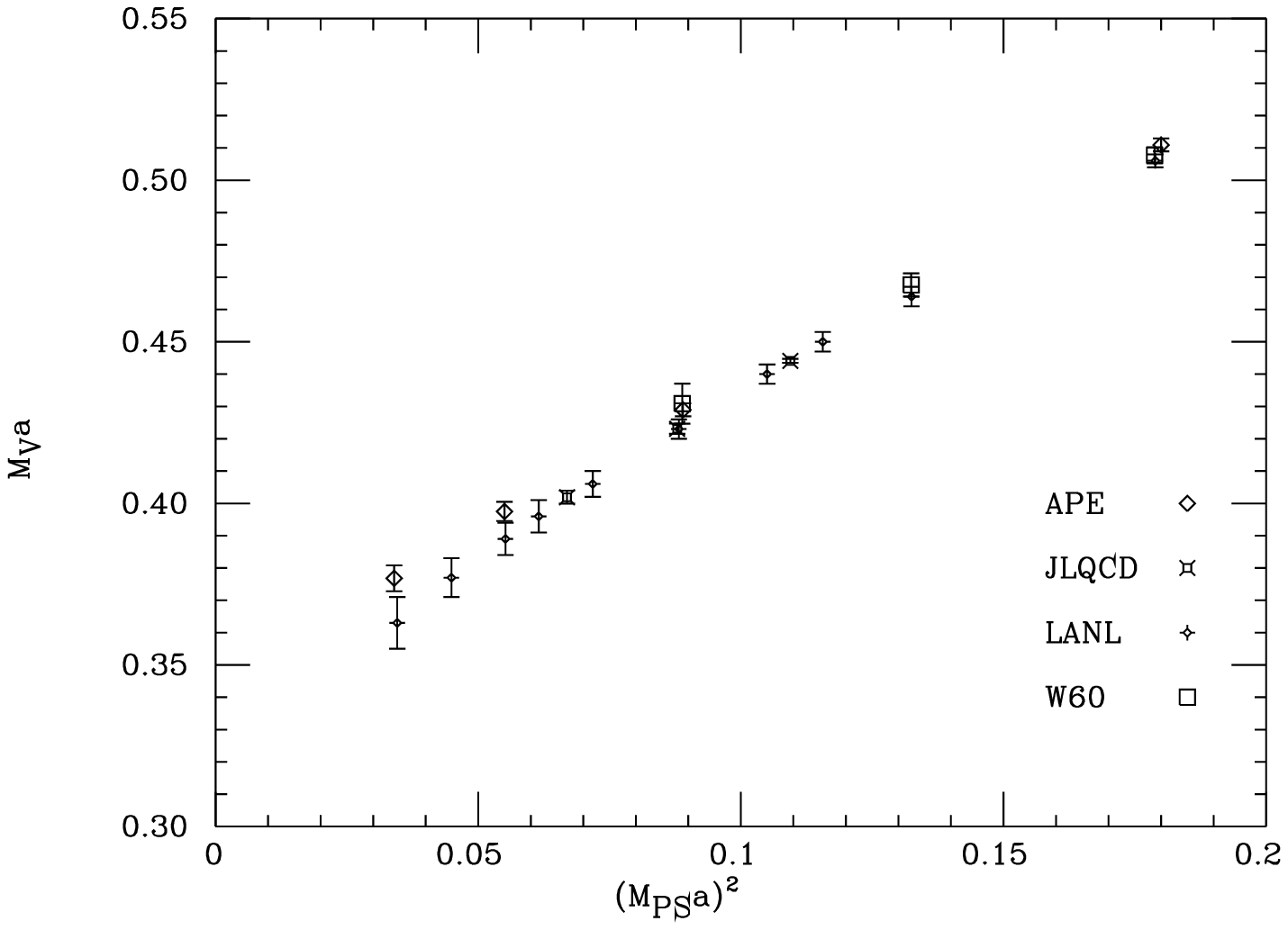}
\caption[]{Vector meson masses in lattice units at $\beta=6.0$ for Wilson action
from Lattice W60, APE $24^3 \times 32$ \cite{parisi}, JLQCD \cite{fukugita} and 
LANL \cite{gupta}.}
\protect\label{fig:vol_60}
\end{figure}

\begin{figure}[htb] 
\putfig{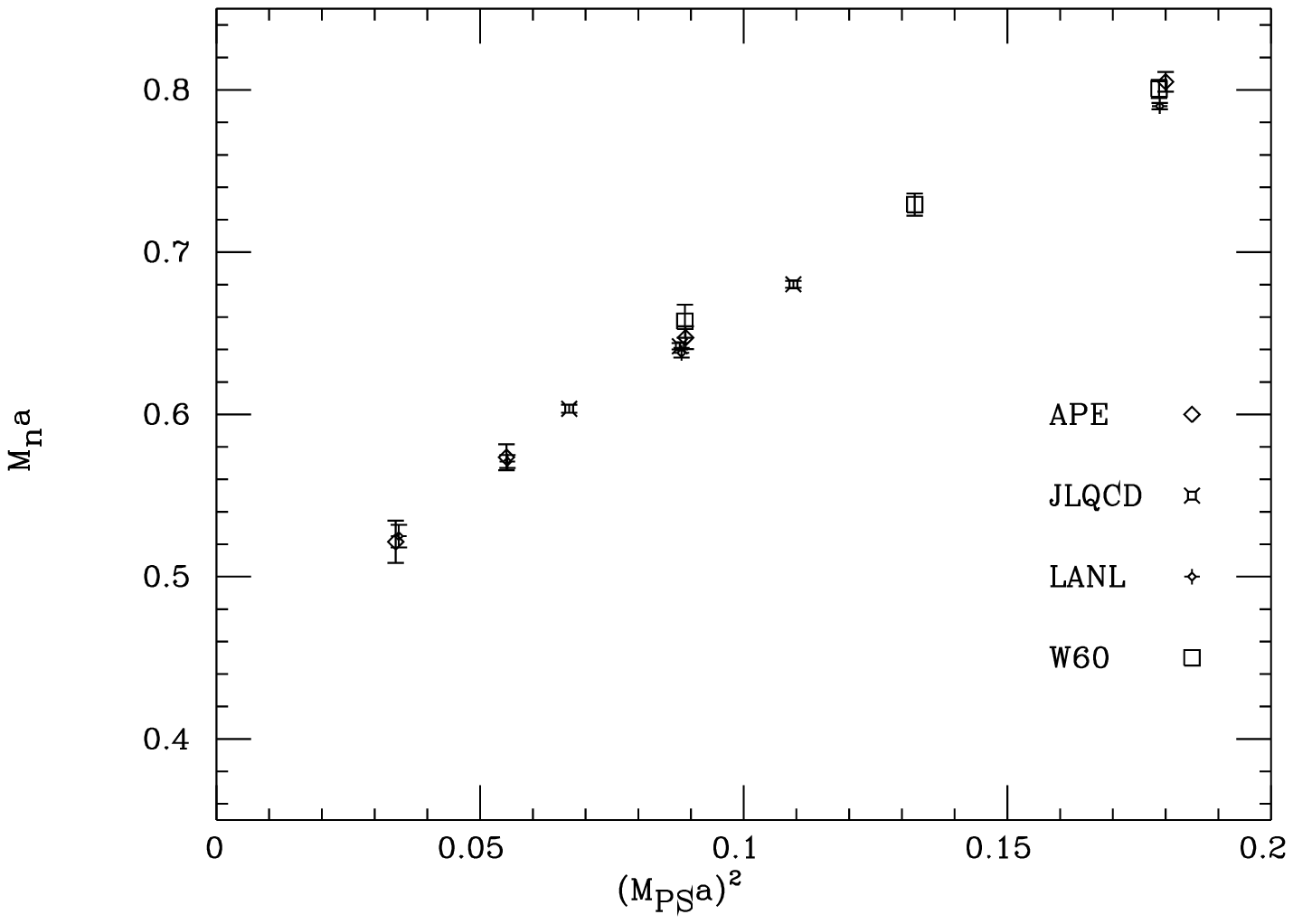}
\caption[]{Nucleon masses in lattice units at $\beta=6.0$ for wilson action
from Lattice W60, APE $24^3 \times 32$ \cite{parisi}, JLQCD \cite{fukugita} 
and LANL \cite{gupta}.}
\protect\label{fig:baryword}
\end{figure}

\begin{figure}[htb] 
\putfig{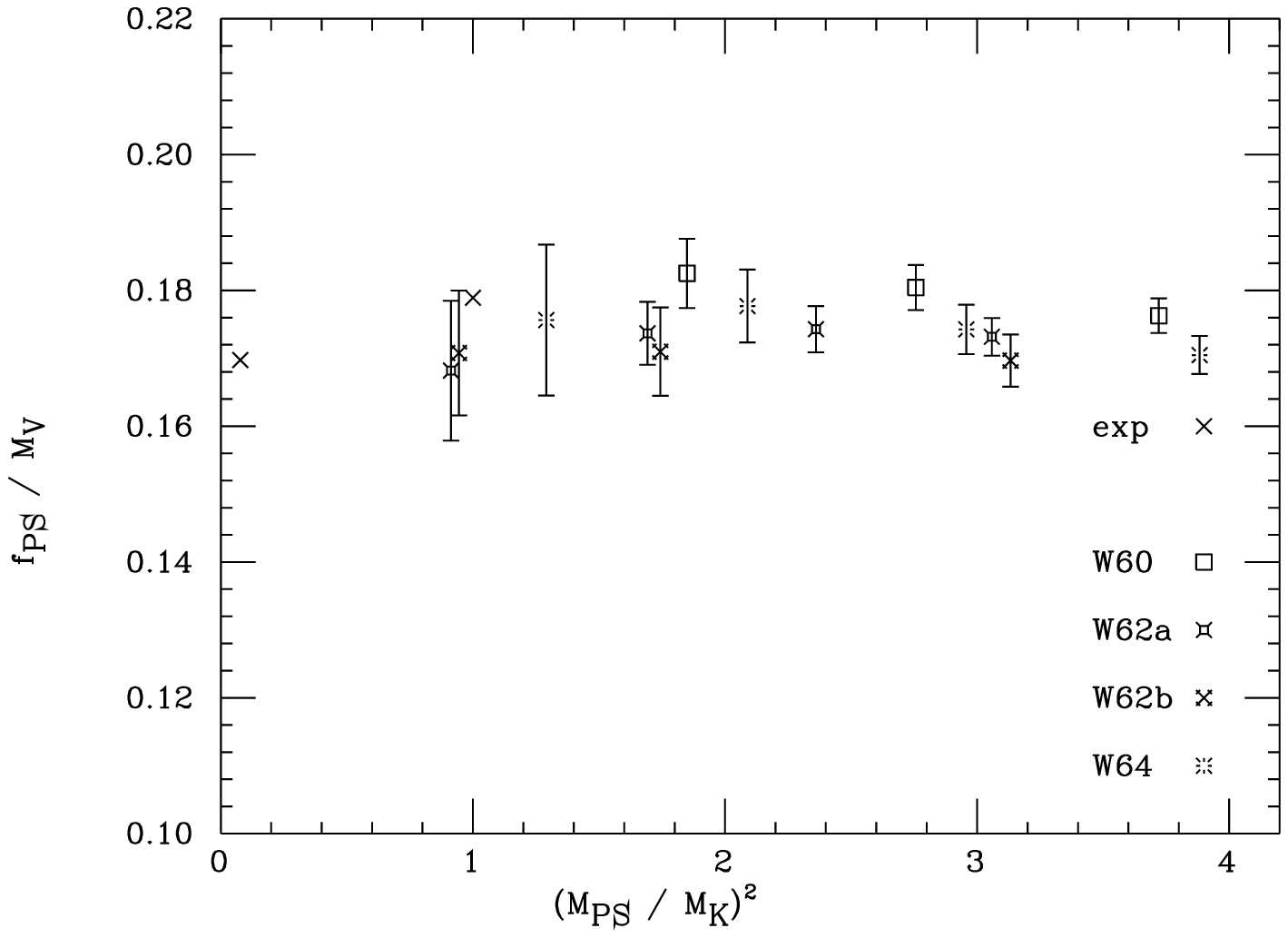}
\caption[]{Values of the ratio $\displaystyle\frac{f_{PS}}{M_V}$ versus the 
squared pseudoscalar mass $M_{PS}^2$ normalized to the lattice value of $M_K$ 
for all lattices with Wilson action.}
\protect\label{fig:fpsmv_mp2wil}
\end{figure}

\begin{figure}[htb] 
\putfig{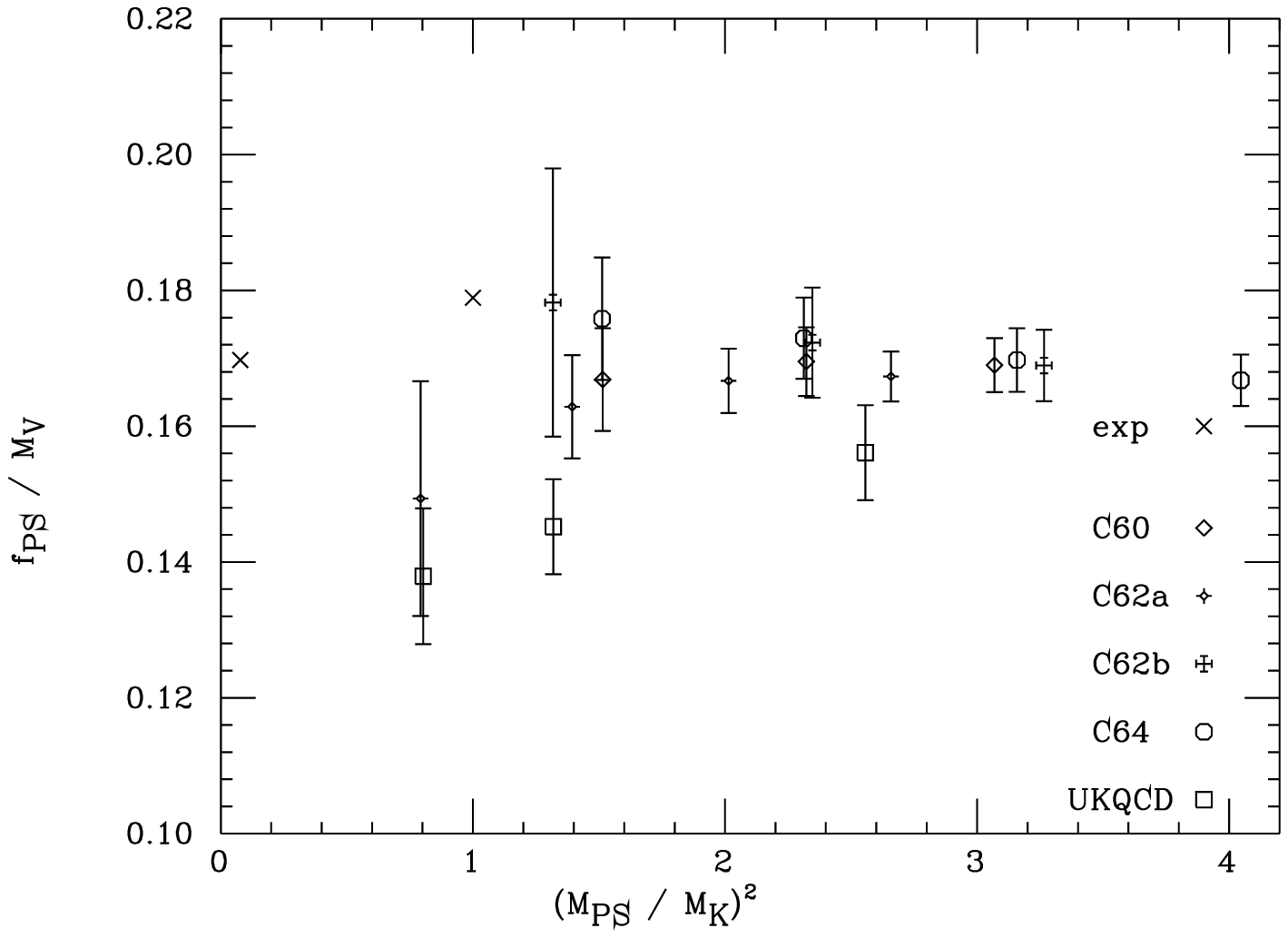}
\caption[]{Values of the ratio $\displaystyle\frac{f_{PS}}{M_V}$ versus the
squared pseudoscalar mass $M_{PS}^2$  normalized to the lattice value of $M_K$
for all lattices with SW-Clover action.}
\protect\label{fig:fpsmv_mp2clo}
\end{figure}

\begin{figure}[htb] 
\putfig{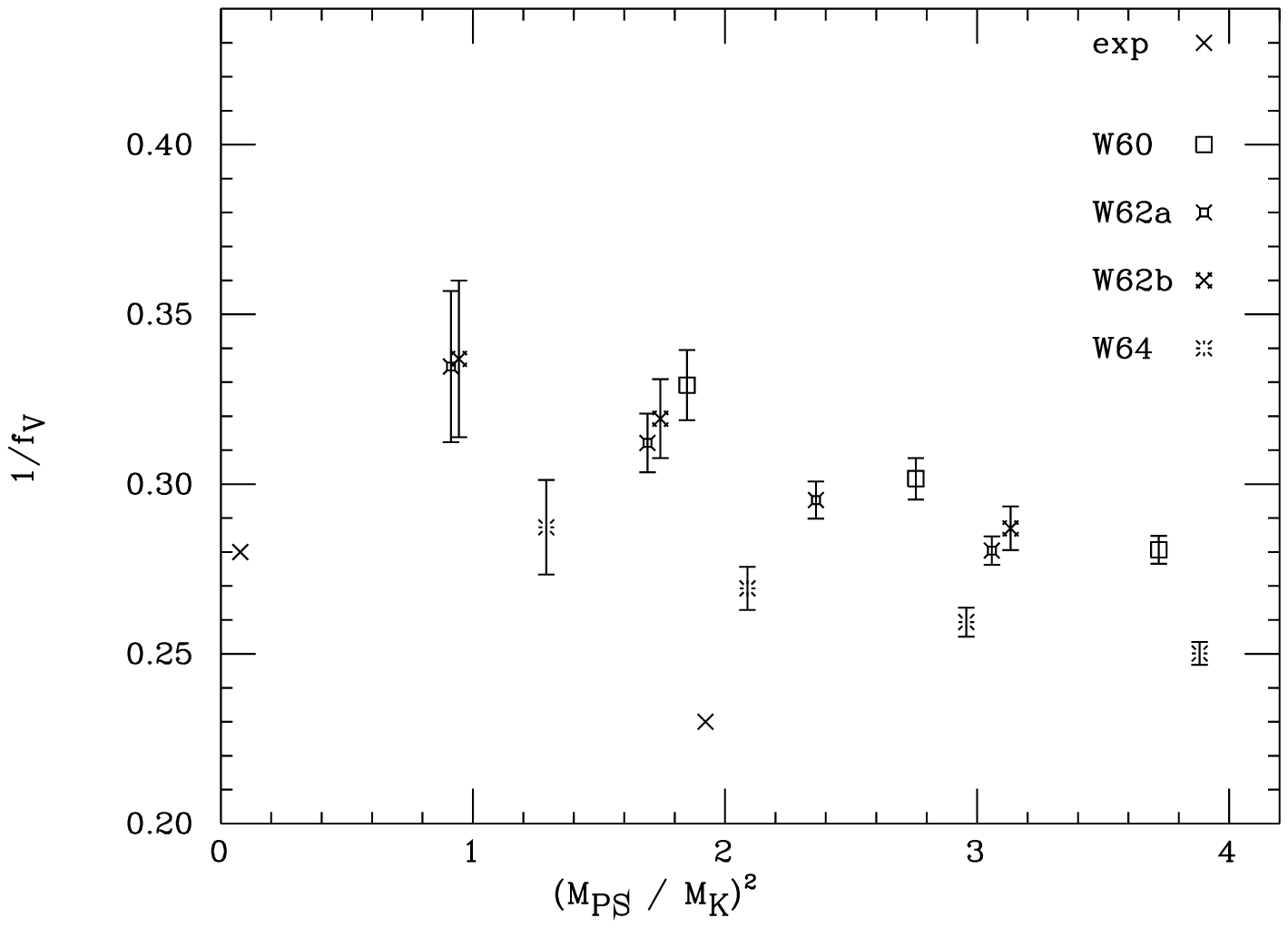}
\caption[]{Values of the vector decay constant $1/f_V$ versus the squared pseudoscalar mass $M_{PS}^2$ 
normalized to the lattice value of $M_K$ for all lattices with Wilson action.}
\protect\label{fig:fvmp2wil}
\end{figure}

\begin{figure}[htb] 
\putfig{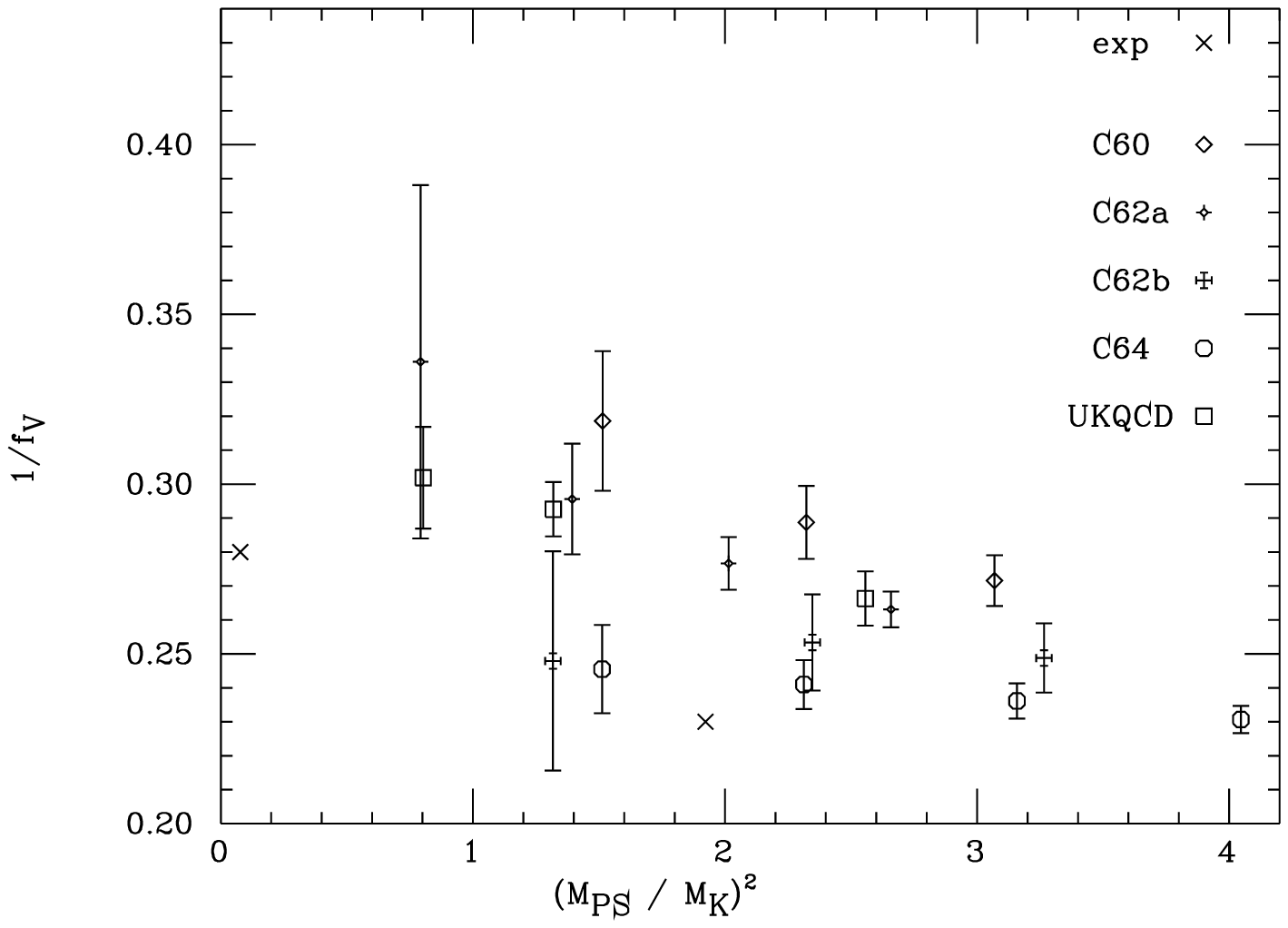}
\caption[]{Values of the vector decay constant $1/f_V$ versus the squared pseudoscalar mass $M_{PS}^2$ 
normalized to the lattice value of $M_K$ for all lattices with SW-Clover action.}
\protect\label{fig:fvmp2clo}
\end{figure}

\begin{figure}[htb] 
\putfig{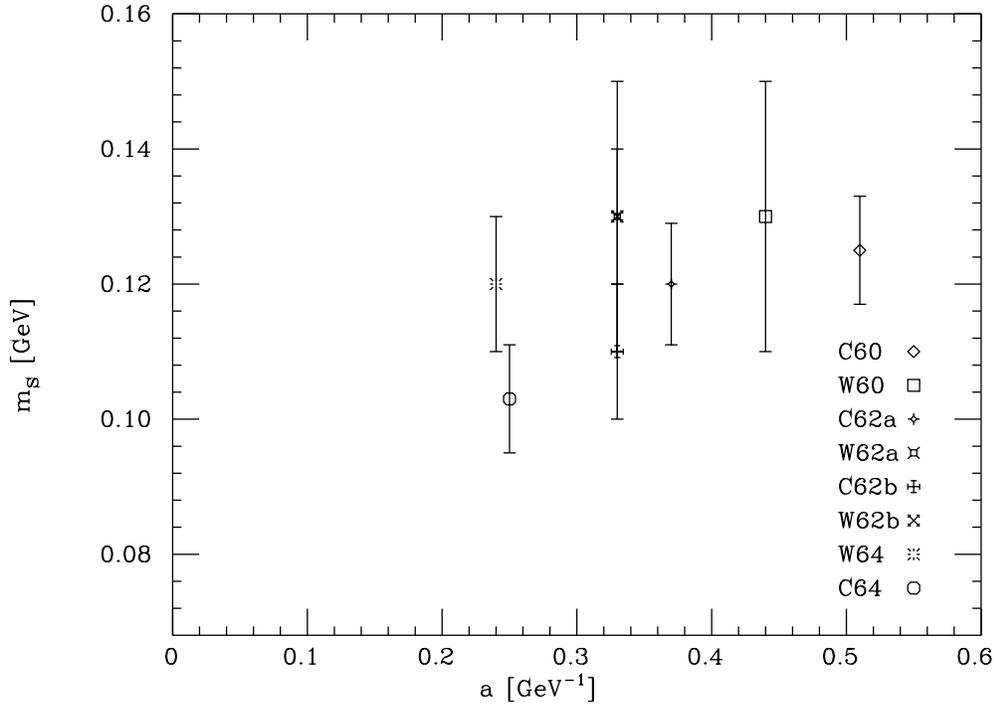}
\caption[]{Renormalized quark masses $m_s^{\overline{MS}}$ for all lattices.}
\protect\label{fig:qmas}
\end{figure}

In this section we analyze the effect of the quenched approximation using the 
$J$ variables first proposed for mesons by Lacock and Michael \cite{michael}.\\
In the spirit of the Maiani-Martinelli proposal \cite{mm}\cite{ukqcd93} to use
$(M_{K^*}-M_{\rho})/(M^2_{K}-M^2_{\pi})$ as an independent way to determine the
lattice scale and avoiding the chiral extrapolation as described in section
\ref{new_method}, Lacock and Michael proposed that
\be
J = M_{K^*}\frac{dM_V}{dM^2_{PS}},
\ee
be computed on the lattice and compared with experimental values.
$J$ is a combination of parameters which is independent of $a$ and $K$
and which does not require any chiral extrapolation to the chiral
limit. We use the experimental values of the meson masses reported in
Table~\ref{tab:mesonsGeV} to determine the experimental values of  $J$. 
This gives the dimensionless value reported in Table~\ref{tab:J}.
\\
Using the method described in section \ref{new_method} it is very easy to
extract the $J$ value for mesons. The results we have
obtained are reported in the Table~\ref{tab:J}. The lattice
values for $J$ are significantly different to the experimental
value and one can see a dependence on $\beta$ but as discussed before a volume
effect may be present on C62b, W64 and C64 which prevents a clear 
conclusion. The disagreement between the data and the experimental value of $J$
is also evident in fig.~\ref{fig:mv_mp2}. It is easy to see that, in the  plane
of  fig.~\ref{fig:mv_mp2}, the $J$ value is proportional to the slope of the
line through the experimental data which is clearly different to the slope of
the lattice data.

As well as defining the quantity $J$ for the meson sector, a
corresponding quantity for the nucleon and delta sector can be
introduced. This can be viewed as the slope of the data in
figs.~\ref{fig:mb_mp2},~\ref{fig:mdelta_mp2}.
While the data is noisy, it is likely that the lattice data also
underestimates the slope of the experimental points in these plots.

\section{Conclusions}
We have analyzed a large set of data on lattices of different lattice
spacing and lattice sizes with both the Wilson and SW-Clover actions.
These simulations were not originally performed for a hadron spectrum
study, hence the relatively large values for the light quark masses.
To overcome potential problems with the chiral extrapolation, a new
technique of obtaining lattice predictions without any
extrapolation in the quark mass was developed.

In the $\beta$ range studied, we conclude that there is no statistical
evidence for an $a$ dependence in the hadron masses
or for the pseudoscalar decay constants.
Our data agrees with the experimental data to within
$\sim 5\%$ for mesonic masses and $\sim 10\%~-~15\%$ for baryonic
masses and the pseudoscalar decay constants. A larger deviation is
present for the vector decay constants which, in our opinion,
deserve a more careful study at larger volumes and $\beta$.
The effects of finite size, if present, are within the statistical
errors of our data.

We obtain for the strange quark mass $m_s^{\overline{MS}}(\mu = 2 GeV)
= 122 \pm 20 \; MeV$ without seeing any clear $a$ dependence.
The meson dispersion relations and the quantity $J$ were also studied.

We believe that the lattice results at $\beta \ge 6$ are quite stable
and show a scaling behaviour, within current statistical
errors.

\begin{ack} 
We wish to thank the APE collaboration for allowing us to use the
lattice correlation functions presented here.
We warmly thank R.~Gupta, V.~Lubicz, G.~Martinelli, G.~Parisi, S.~Sharpe and
A.~Vladikas for useful discussions.
We also thank E.~Franco for providing us the code for the computation of 
the renormalization constants of vector and axial currents.
This work was partially supported by the EC Contract `Computational 
Particle Physics' CHRX-CT92-0051, and by the EC Human Capital 
and Mobility Program, contracts ERBCHBICT941462 and ERBCHBGCT940665.
V.G.~wishes to thank the Dipartimento di Fisica ``G. Marconi''
of the Universit\`a di Roma ``La Sapienza'' for its hospitality,
and acknowledges partial support by CICYT under grant number 
AEN-96/1718.
\end{ack}



\newpage
\clearpage
\setlength{\tabcolsep}{.6pc}
\begin{table}
\caption{Hadron masses and meson decay constants versus the hopping 
parameter for Lattice C60.}
\label{tab:latI}

\begin{tabular}{ccccccc}
\hline
K&$M_{PS}a$&$M_Va$&$M_na$&$M_{\delta}a$&$f_{PS}a/Z_A$&$1/(f_V Z_V)$\\
\hline
0.1425&0.439(1)&0.549(5)&0.833(9)&0.90(2)&0.095(2)&0.324(9)\\
0.1432&0.382(1)&0.512(7)&0.76(1)&0.84(3)&0.089(2)&0.34(1)\\
0.1440&0.308(1)&0.48(1)&0.69(1)&0.78(5)&0.082(3)&0.38(2)\\
\hline
0.14549(2)& - &0.40(2)&0.54(2)&0.65(7)&0.069(4)&0.43(4)\\ 
\hline
\hline
\end{tabular}
\end{table}
\setlength{\tabcolsep}{.6pc}
\begin{table}

\caption{As in Table~{\protect \ref{tab:latI}} for Lattice W60.}
\label{tab:latII}

\begin{tabular}{ccccccc}
\hline
K&$M_{PS}a$&$M_Va$&$M_na$&$M_{\delta}a$&$f_{PS}a/Z_A$&$1/(f_V Z_V)$\\
\hline
0.1530&0.423(1)&0.508(3)&0.801(6)&0.864(8)&0.114(2)&0.391(6)\\
0.1540&0.364(1)&0.468(4)&0.729(7)&0.81(1) &0.108(2)&0.420(8)\\
0.1550&0.298(1)&0.431(6)&0.66(1) &0.75(2) &0.100(2)&0.46(1) \\
\hline
0.15703(2)& - &0.353(9)&0.52(1)&0.64(3)&0.087(3)&0.52(2)\\
\hline
\hline
\end{tabular}
\end{table}
\setlength{\tabcolsep}{.6pc}
\begin{table}

\caption{As in Table~{\protect \ref{tab:latI}} for Lattice C62a.}
\label{tab:latIII}

\begin{tabular}{ccccccc}             
\hline
K&$M_{PS}a$&$M_Va$&$M_na$&$M_{\delta}a$&$f_{PS}a/Z_A$&$1/(f_V Z_V)$\\
\hline
0.14144&0.297(1)&0.386(3)&0.586(5)& - &0.066(1)&0.310(6)\\
0.14184&0.258(1)&0.362(5)&0.542(8)& - &0.062(2)&0.326(9)\\
0.14224&0.215(1)&0.34(1) &0.50(1) & - &0.057(2)&0.35(2)\\
0.14264&0.162(2)&0.34(3) &0.46(4) & - &0.052(3)&0.40(6)\\
\hline
0.14315(1)& -   &0.29(2) &0.40(3)& - &0.046(3)&0.39(3)\\
\hline
\hline
\end{tabular}
\end{table}
\setlength{\tabcolsep}{.6pc}
\begin{table}

\caption{As in Table~{\protect \ref{tab:latI}} for Lattice W62a.}
\label{tab:latIV}

\begin{tabular}{ccccccc}
\hline
K&$M_{PS}a$&$M_Va$&$M_na$&$M_{\delta}a$&$f_{PS}a/Z_A$&$1/(f_V Z_V)$\\
\hline
0.1510&0.289(1)&0.366(2)&0.566(4)& - &0.080(1)&0.381(6)\\
0.1515&0.254(1)&0.343(3)&0.525(6)& - &0.075(1)&0.401(7)\\
0.1520&0.215(1)&0.321(5)&0.48(1) & - &0.070(2)&0.42(1)\\
0.1526&0.158(1)&0.29(1) &0.45(3) & - &0.062(2)&0.45(3)\\
\hline
0.15329(1)& -  &0.26(1) &0.38(2)& - &0.056(2)&0.48(2)\\
\hline
\hline
\end{tabular}
\end{table}
\setlength{\tabcolsep}{.6pc}
\begin{table}

\caption{As in Table~{\protect \ref{tab:latI}} for Lattice C62b.}
\label{tab:latV}

\begin{tabular}{ccccccc}
\hline
K&$M_{PS}a$&$M_Va$&$M_na$&$M_{\delta}a$&$f_{PS}a/Z_A$&$1/(f_V Z_V)$\\
\hline
0.14144&0.294(3)&0.378(6)&0.59(2)&0.68(1)&0.066(2)&0.29(1)\\
0.14190&0.249(3)&0.34(1) &0.53(2)&0.65(2)&0.061(2)&0.30(2)\\
0.14244&0.186(4)&0.30(3) &0.43(5)&0.57(4)&0.055(3)&0.29(4)\\
\hline
0.14312(4)& -   &0.25(3) &0.35(7)&0.53(6)&0.048(4)&0.30(4)\\
\hline
\hline
\end{tabular}
\end{table}
\setlength{\tabcolsep}{.6pc}
\begin{table}

\caption{As in Table~{\protect \ref{tab:latI}} for Lattice W62b.}
\label{tab:latVI}

\begin{tabular}{ccccccc}
\hline
K&$M_{PS}a$&$M_Va$&$M_na$&$M_{\delta}a$&$f_{PS}a/Z_A$&$1/(f_V Z_V)$\\
\hline
0.1510&0.291(1)&0.370(3)&0.576(7)&0.618(7)&0.079(2)&0.390(9)\\
0.1520&0.217(1)&0.323(7)&0.50(2) &0.56(2) &0.069(2)&0.43(2)\\
0.1526&0.160(2)&0.29(1) &0.42(6) &0.55(4) &0.062(2)&0.46(3)\\
\hline
0.15330(2)& -  &0.26(1) &0.38(6) &0.49(4)&0.055(2)&0.49(3)\\
\hline
\hline
\end{tabular}
\end{table}
\setlength{\tabcolsep}{.6pc}
\begin{table}

\caption{As in Table~{\protect \ref{tab:latI}} for Lattice W64.}
\label{tab:latVII}

\begin{tabular}{ccccccc}
\hline
K&$M_{PS}a$&$M_Va$&$M_na$&$M_{\delta}a$&$f_{PS}a/Z_A$&$1/(f_V Z_V)$\\
\hline
0.1488&0.235(1)&0.289(2)&0.461(4)&0.499(7)&0.0609(9)&0.333(4)\\
0.1492&0.205(2)&0.266(3)&0.423(5)&0.47(1)&0.0573(9)&0.346(6)\\
0.1496&0.172(2)&0.242(4)&0.38(1)&0.42(2)&0.053(1)&0.359(8)\\
0.1500&0.136(3)&0.22(1)&0.33(3)&0.36(4)&0.048(2)&0.38(2)\\
\hline
0.15058(3)& - &0.188(9)&0.28(2)&0.33(4)&0.043(2)&0.39(2)\\
\hline
\hline
\end{tabular}
\end{table}
\setlength{\tabcolsep}{.6pc}
\begin{table}

\caption{As in Table~{\protect \ref{tab:latI}} for Lattice C64.}
\label{tab:latVIII}

\begin{tabular}{ccccccc}
\hline
K&$M_{PS}a$&$M_Va$&$M_na$&$M_{\delta}a$&$f_{PS}a/Z_A$&$1/(f_V Z_V)$\\
\hline
0.1400&0.249(1)&0.306(2)&0.484(4)&0.52(1)&0.0524(9)&0.269(5)\\
0.1403&0.220(1)&0.283(3)&0.446(6)&0.49(1)&0.049(1)&0.276(6)\\
0.1406&0.188(2)&0.259(5)&0.40(1)&0.44(2)&0.046(1)&0.281(8)\\
0.1409&0.152(3)&0.23(1)&0.34(2)&0.37(4)&0.042(1)&0.29(2)\\
\hline
0.14143(3)& - &0.19(1)&0.29(2)&0.32(5)&0.037(2)&0.30(2)\\
\hline
\hline
\end{tabular}
\end{table}
\newpage
\clearpage

\setlength{\tabcolsep}{.5pc}
\begin{table}

\caption{Values of $a^{-1}$ in $GeV$ from different observables for 
all lattices. The values of $a^{-1}_{K^*}$ (shown in boldface) are taken as our
preferred estimate of the inverse lattice spacing.}
\label{tab:inva}

\begin{tabular}{lcccccccc}
\hline
&C60&W60&C62a&W62a &C62b&W62b&W64&C64\\
\hline
$a^{-1}_\rho   $&1.88(9)  &2.15(5)&  2.6(2)& 2.9(1)&3.0(3)&2.9(2)&4.0(2)&3.9(2)\\
{\bf $a^{-1}_{K^*}$}& {\bf 1.98(8)}  &{\bf 2.26(5)}&{\bf  2.7(1)}& {\bf
3.00(9)}&
{\bf 3.0(3)}&{\bf 3.0(1)}&{\bf 4.1(2)}&{\bf 4.0(2)}\\
$a^{-1}_{\eta_{ss}}$&1.98(8)  &2.26(5)&  2.7(1)&3.00(9)&3.0(3)&3.0(1)&4.2(2)&4.0(2)\\
$a^{-1}_\phi   $&2.06(7)  &2.35(5)&  2.8(1)&3.10(8)&3.1(3)&3.1(1)&4.3(2)&4.1(2)\\
$a^{-1}_N      $&1.70(6)  &1.79(5)&  2.3(1)& 2.4(1)&2.7(5)&2.4(3)&3.2(2)&3.2(2)\\
$a^{-1}_{\Lambda\Sigma} $&1.86(6)  &1.99(4)& 2.5(1) &2.7(1)&2.8(4)&2.7(3)&
3.6(2)&3.5(2)\\
$a^{-1}_{\Xi}  $&1.98(6)  &2.13(4)& 2.68(8)& 2.84(8)&2.9(4)&2.8(2)&3.8(2)&3.7(2)\\
$a^{-1}_\Delta $&1.9(2)   &1.90(8)& -      &  -    &2.3(2)&2.5(2)&3.7(4)&3.8(5)\\
$a^{-1}_\Omega $&2.07(9)  &2.21(6)& -      &  -    &2.7(1)&2.9(1)&4.0(2)&4.0(3)\\
$a^{-1}_{f_\pi}$&1.9(1)   &1.90(7)&  2.9(2)& 2.9(1)&2.8(2)&2.9(1)&3.7(2)&3.6(2)\\
$a^{-1}_{f_K}$  &2.11(8)  &2.16(6)&  3.0(1)& 3.1(1)&3.1(2)&3.17(9)&4.1(1)& 4.0(1)\\
\hline
\hline
\end{tabular}
\end{table}

\setlength{\tabcolsep}{1.5pc}
\begin{table}

\caption{Extrapolated/interpolated meson masses in lattice units 
for all lattices.}
\label{tab:mesons}

\begin{tabular}{lccccc}
\hline
& $M_\rho a$ & $M_{K^*} a$ & $M_{\eta_{ss}} a$ & $M_\phi a$ & \\
\hline
C60   &   0.41(2)  &   0.45(2)    & 0.35(1)  &   0.49(2)  &  \\
W60  &   0.358(9) &   0.395(9)   & 0.303(7) &   0.433(9) &  \\
C62a &   0.30(2)  &   0.33(2)    & 0.25(1)  &   0.36(1)  &  \\
W62a  &   0.27(1)  &   0.298(9)   & 0.229(7) &   0.329(9) &  \\
C62b   &   0.26(3)  &   0.29(3)    & 0.23(2)  &   0.33(3)  &  \\
W62b  &   0.26(1)  &   0.30(1)    & 0.23(1)  &   0.33(1)  &  \\
W64 &   0.192(8) &   0.215(8)   & 0.165(6) &   0.239(8) &  \\
C64&   0.20(1)  &   0.22(1)    & 0.171(8) &   0.25(1)  &  \\
\hline
\hline
\end{tabular}
\end{table}
\setlength{\tabcolsep}{.6pc}
\begin{table}

\caption{Extrapolated/interpolated meson decay constants for all lattices.}
\label{tab:decay}

\begin{tabular}{lcccccc}
\hline
& $f_\pi a/(Z_A m_\rho a )$ & $1/(f_\rho Z_V)$ & $f_K a/(Z_A m_{K^*} a )$ & 
$1/(f_{K^*} Z_V)$ & $1/(f_{\phi} Z_V)$ & $f_K/f_\pi$\\
\hline
C60   &   0.17(1)  &   0.42(3)  &   0.172(9)  &  0.39(2) & 0.36(1) & 1.11(2)\\
W60  &   0.25(1)  &   0.51(2)  &   0.239(8)  &  0.48(2) & 0.45(1) & 1.08(1)\\
C62a &   0.16(1)  &   0.39(3)  &   0.164(9)  &  0.36(2) & 0.332(8)& 1.15(3)\\
W62a  &   0.21(1)  &   0.47(2)  &   0.214(8)  &  0.45(1) & 0.417(8)& 1.13(1)\\
C62b   &   0.19(3)  &   0.30(4)  &   0.18(2)   &  0.30(3) & 0.30(2) & 1.10(3)\\
W62b  &   0.21(1)  &   0.49(3)  &   0.21(1)   &  0.46(2) & 0.43(1) & 1.12(2)\\
W64 &   0.23(2)  &   0.39(2)  &   0.22(1)   &  0.38(1) & 0.363(8)& 1.10(2)\\
C64&   0.19(1)  &   0.30(2)  &   0.18(1)   &  0.29(1) & 0.284(9)& 1.10(1)\\
\hline                                                                     
\hline
\end{tabular}
\end{table}
\setlength{\tabcolsep}{1.2pc}
\begin{table}

\caption{Extrapolated/interpolated baryon masses in lattice units  
for all lattices.}
\label{tab:baryons}

\begin{tabular}{lcccccc}
\hline
& $M_N a$ & $M_{\Lambda\Sigma} a$ & $M_\Xi a$& $M_\Delta a$ & $M_\Omega a$ &\\
\hline
C60   &   0.55(2)  &  0.61(2)  &  0.67(2)  & 0.66(6)  &   0.81(3)  &    \\
W60  &   0.53(1)  &  0.57(1)  &  0.62(1)  & 0.65(3)  &   0.76(2)  &    \\
C62a &   0.41(3)  &  0.45(2)  &  0.49(1)  & -        &   -        &    \\
W62a  &   0.39(2)  &  0.43(2)  &  0.46(1)  & -        &   -        &    \\
C62b   &   0.35(7)  &  0.40(6)  &  0.45(5)  & 0.53(6)  &   0.62(2)  &    \\
W62b  &   0.39(5)  &  0.43(5)  &  0.46(4)  & 0.50(4)  &   0.57(2)  &    \\
W64 &   0.29(2)  &  0.32(2)  &  0.35(1)  & 0.34(4)  &   0.42(2)  &    \\
C64&   0.29(2)  &  0.32(2)  &  0.35(2)  & 0.33(5)  &   0.42(3)  &    \\
\hline
\hline
\end{tabular}
\end{table}
\newpage
\clearpage

\begin{table}
\caption{Predicted meson masses in $GeV$ for all lattices, using  the scale from 
$M_{K^*}$.}
\label{tab:mesonsGeV}
\begin{tabular}{lccc}
\hline
& $M_\rho$ & $M_{\eta_{ss}}$ & $M_\phi$ \\
\hline
Exper.     & 0.770    & 0.686 (see text)   & 1.019   \\
\hline
C60   & 0.809(7) & 0.6849(3) & 0.977(7) \\
W60  & 0.808(3) & 0.6849(1) & 0.978(3) \\
C62a & 0.81(1)  & 0.6849(5) & 0.98(1) \\
W62a  & 0.803(6) & 0.6851(2) & 0.984(6) \\
C62b   & 0.79(1)  & 0.6856(5) & 1.00(1) \\
W62b  & 0.797(7) & 0.6853(3) & 0.989(7)\\
W64 & 0.796(4) & 0.6853(2) & 0.990(4)\\
C64& 0.792(4) & 0.6855(2) & 0.994(4)\\
\hline
\hline
\end{tabular}
\end{table}
\setlength{\tabcolsep}{1.2pc}
\begin{table}
\caption{Predicted baryon masses in $GeV$ for all lattices, using  the scale from 
$M_{K^*}$.}
\label{tab:baryonsGeV}
\begin{tabular}{lccccc}
\hline
& $M_N $ & $M_{\Lambda\Sigma}$ & $M_\Xi $& $M_\Delta $ & $M_\Omega $\\
\hline
Exper.  & 0.9389  & 1.135   & 1.3181  & 1.232   & 1.6724 \\
\hline
C60   & 1.09(5) & 1.21(4) & 1.32(4) & 1.3(1)  & 1.60(9) \\   
W60  & 1.19(5) & 1.29(4) & 1.40(4) & 1.46(7) & 1.71(4) \\   
C62a & 1.1(1)  & 1.22(8) & 1.34(7) &  -      &  -   \\   
W62a  & 1.17(7) & 1.28(6) & 1.39(5) &  -      &  -   \\   
C62b   & 1.1(2)  & 1.2(2)  & 1.4(1)  & 1.6(3)  & 1.9(2) \\   
W62b  & 1.2(1)  & 1.3(1)  & 1.40(9) & 1.50(9) & 1.72(5) \\   
W64 & 1.21(9) & 1.32(8) & 1.43(6) & 1.4(2)  & 1.72(9) \\   
C64& 1.2(1)  & 1.29(8) & 1.41(7) & 1.3(2)  & 1.7(1) \\   
\hline
\hline
\end{tabular}
\end{table}
\setlength{\tabcolsep}{1.2pc}
\begin{table}
\caption{Predicted meson decay constants for all lattices, 
using  the scale from $M_{K^*}$.}
\label{tab:decaysGeV}
\begin{tabular}{lccccc}
\hline
&$f_\pi$~(GeV)& 
$\displaystyle\frac{1}{f_\rho}$&$f_K$~($GeV$) 
&$\displaystyle\frac{1}{f_{K^*}}$& $\displaystyle\frac{1}{f_{\phi}}$\\
\hline
Exper.  & 0.1307   & 0.28  & 0.1598   &   & 0.23\\
\hline
C60   & 0.134(9) & 0.35(3) & 0.149(8) & 0.33(2) & 0.30(1)\\ 
W60  & 0.155(7) & 0.37(2) & 0.167(6) & 0.35(1) & 0.324(8)\\ 
C62a & 0.124(9) & 0.33(3) & 0.143(8) & 0.30(2) & 0.281(7)\\ 
W62a  & 0.135(6) & 0.35(2) & 0.153(5) & 0.33(1) & 0.307(6)\\ 
C62b   & 0.14(2)  & 0.26(4) & 0.16(2)  & 0.25(3) & 0.25(2)\\ 
W62b  & 0.135(8) & 0.36(2) & 0.152(7) & 0.34(2) & 0.315(9)\\ 
W64 & 0.147(9)  & 0.29(1) & 0.161(8) & 0.283(9) & 0.272(6)\\ 
C64& 0.144(9) & 0.25(1) & 0.158(8) & 0.25(1) & 0.243(8)\\ 
\hline
\hline
\end{tabular}
\end{table}
\newpage
\clearpage

\setlength{\tabcolsep}{.4pc}
\begin{table}
\caption{Values of $E(\vec{p})$ for mesons with non-zero momentum versus 
the hopping parameter.}
\label{tab:disp}
\begin{tabular}{lcccc|cccc}
\hline
& $K$ &            $p^2=1$& $p^2=2$ & $p^2=3$  &$p^2=1$& $p^2=2$ & $p^2=3$ & \\
\hline
C60   & 0.1425  & 0.567(4) & 0.67(1) & 0.73(3)  & 0.654(4) & 0.751(8)& 0.82(1) & \\ 
        & 0.1432  & 0.526(6) & 0.63(2) & 0.68(4)  & 0.625(6) & 0.73(1) & 0.78(1) & \\ 
        & 0.1440  & 0.479(9) & 0.57(4) & 0.60(8)  & 0.60(1)  & 0.70(2) & 0.70(2) & \\ 
\hline
W60$ $& 0.1530& 0.554(5) & 0.66(1) & 0.73(2)  & 0.621(4) & 0.716(7)& 0.79(1) & \\
        & 0.1540  & 0.513(7) & 0.62(2) & 0.68(4)  & 0.591(6) & 0.692(9)& 0.75(2) & \\ 
        & 0.1550  & 0.47(1)  & 0.58(3) & 0.62(7)  & 0.566(8) & 0.67(2) & 0.69(3) & \\ 
\hline
C62a & 0.14144 & 0.399(3) & 0.479(6) & 0.53(2) & 0.476(3) & 0.557(4) & 
0.618(8) & \\
        & 0.14184 & 0.371(4) & 0.455(9) & 0.51(4) & 0.455(3) & 0.543(7) & 
0.60(1) & \\ 
        & 0.14224 & 0.340(7) & 0.43(1)  & 0.50(7) & 0.436(5) & 0.54(1)  & 
0.59(2) & \\ 
        & 0.14264 & 0.30(1)  & 0.40(3)  &    -    & 0.42(1)  & 0.56(3)  & 
0.55(3) & \\ 
\hline
W62a  & 0.1510  & 0.395(3) & 0.479(6) & 0.52(2) & 0.459(2) & 0.541(3) & 
0.605(7) & \\
        & 0.1515  & 0.369(3) & 0.459(9) & 0.50(3) & 0.440(3) & 0.528(5) & 
0.592(9) & \\ 
        & 0.1520  & 0.343(5) & 0.44(1)  & 0.48(5) & 0.423(3) & 0.518(8) & 
0.58(1) & \\ 
        & 0.1526  & 0.31(1)  & 0.41(3)  & 0.5(1)  & 0.404(6) & 0.53(2)  & 
0.54(3) & \\ 
\hline
C62b   & 0.14144 & 0.48(1)  & 0.60(3) &  -       & 0.536(7) & 0.66(2) & 0.77(4) & \\
        & 0.14190 & 0.46(2)  & 0.56(4) &  -       & 0.53(1)  & 0.64(3) & 0.77(5) & \\ 
        & 0.14244 & 0.42(6)  & 0.48(7) &  -       & 0.53(3)  & 0.63(5) & 0.76(8) & \\ 
\hline
W62b  & - & - & - & - & - & - & - & \\
&&&&&&&&\\ 
\hline
W64 & 0.1488  & 0.363(6) & 0.47(2) & -    & 0.406(4) & 0.50(1) & - & \\
        & 0.1492  & 0.35(1)  & 0.46(4) & -    & 0.394(5) & 0.50(2) & - & \\ 
        & 0.1496  & 0.33(2)  & 0.46(8) & -    & 0.385(8) & 0.51(4) & - & \\ 
        & 0.1500  & 0.33(5)  & 0.4(2) & -     & 0.38(2)  & 0.6(1)  & - & \\ 
\hline
C64& 0.1400  & 0.371(6) & 0.49(2) & -    & 0.419(4) & 0.51(1) & - & \\
        & 0.1403  & 0.36(1)  & 0.49(4) & -    & 0.406(5) & 0.51(2) & - & \\ 
        & 0.1406  & 0.34(2)  & 0.50(8) & -    & 0.393(8) & 0.52(5) & - & \\ 
        & 0.1409  & 0.33(4)  & 0.5(2)  & -    & 0.38(1)  & 0.6(2)  & - & \\ 
\hline
\hline
\end{tabular}
\end{table}

\setlength{\tabcolsep}{1.2pc}
\begin{table}

\caption{Values for $K_c$, $K_l$ and $K_s$ for all
lattices.}
\label{tab:kappas}

\begin{tabular}{lccc}
\hline
& $K_c$ & $K_l$ & $K_s$\\
\hline
C60   &  0.14549(2)  &  0.14540(2)  &  0.1436(1)  \\
W60  &  0.15703(2)  &  0.15693(2)  &  0.15492(9) \\
C62a &  0.14315(1)  &  0.14310(2)  &  0.1419(1)  \\
W62a  &  0.15329(1)  &  0.15322(2)  &  0.15184(9) \\
C62b    &  0.14312(4)  &  0.14308(4)  &  0.1421(2) \\
W62b   &  0.15330(2)  &  0.15323(2)  &  0.1519(1) \\
W64  &  0.15058(3)  &  0.15054(3)  &  0.14969(8)\\
C64 &  0.14143(3)  &  0.14140(3)  &  0.14074(7)\\
\hline
\hline
\end{tabular}
\end{table}

\setlength{\tabcolsep}{1.2pc}
\begin{table}

\caption{Values for the lattice strange lattice quark masses for all 
lattices and the corresponding $\overline{MS}$ values at NLO, both in $MeV$. }
\label{tab:quarkm}

\begin{tabular}{lccc}
\hline
       & $m_s(a)$ & $m_s^{\overline{MS}}(\mu=2\;\mbox{$GeV$})$ & \\
\hline
C60    &  89(3)  &  120(10)   &  \\
W60    &  98(2)  &  130(20)   &  \\
C62a   &  83(4)  &  120(10)   &  \\
W62a   &  93(3)  &  130(10)   &  \\
C62b   &  75(6)  &  110(10)   &  \\
W62b   &  92(4)  &  130(20)   &  \\
W64    &  82(3)  &  120(10)   &  \\
C64    &  69(3)  &  100(10)   &  \\
\hline
\hline
\end{tabular}
\end{table}

\setlength{\tabcolsep}{.8pc}
\begin{table}
\caption{Predicted values of $J$ for all lattices.}
\label{tab:J}
\begin{tabular}{lc}
\hline
& $J$ \\
\hline
Exper.  & 0.48 \\
\hline
C60   & 0.34(3)\\ 
W60  & 0.34(1)  \\ 
C62a & 0.34(5) \\ 
W62a  & 0.36(2) \\ 
C62b & 0.42(5) \\ 
W62b  & 0.38(3) \\ 
W64 & 0.39(2)   \\ 
C64& 0.40(2)   \\ 
\hline
\hline
\end{tabular}
\end{table}

\newpage
\clearpage   
\newpage
\clearpage


\newpage
\clearpage

\newpage
\clearpage

\newpage
\clearpage


\end{document}